\RequirePackage{fix-cm}
\documentclass[smallextended]{svjour3}       
\smartqed  
\usepackage{graphicx}
\usepackage{xcolor}
\usepackage{cite}
\usepackage{amsmath,amssymb,amsfonts}
\usepackage{algorithmic}
\usepackage{graphicx}
\usepackage{textcomp}
\usepackage{xcolor}
\usepackage{ifthen}
\usepackage{booktabs}
\usepackage{scalefnt}
\usepackage{color, colortbl}
\usepackage[flushleft]{threeparttable}
\usepackage[square,comma,numbers,sort]{natbib}
\usepackage{balance} 
\usepackage{multirow}
\usepackage{xcolor}
\usepackage{pifont}
\usepackage{subcaption}
\usepackage{caption}
\usepackage{lscape}

\newboolean{showcomments}
\setboolean{showcomments}{true} 
\ifthenelse{\boolean{showcomments}}
  {\newcommand{\nb}[2]{
    \fcolorbox{gray}{yellow}{\bfseries\sffamily\scriptsize#1}
    {\sf\small$\blacktriangleright$\textit{#2}$\blacktriangleleft$}
   }
   
  }
  {\newcommand{\nb}[2]{}
   
  }

\newcommand{\MyBox}[1]{\vspace{4mm}\noindent\framebox[\columnwidth][c]{\parbox[b]{0.95\columnwidth}{ #1 }}\vspace{4mm}}

\newcommand{\MyPara}[1]{\vspace{.2em}\noindent\textit{\textbf{#1}}\hspace{.3em}}

\definecolor{Gray}{gray}{0.9}
\def\BibTeX{{\rm B\kern-.05em{\sc i\kern-.025em b}\kern-.08em
  T\kern-.1667em\lower.7ex\hbox{E}\kern-.125emX}}

\journalname{Empirical Software Engineering manuscript}
\begin{document}

\title{Quality Gatekeepers: Investigating the Effects of Code Review Bots on Pull Request Activities
}

\titlerunning{Quality Gatekeepers}        

\author{Mairieli Wessel         \and
        Alexander Serebrenik    \and
        Igor Wiese              \and
        Igor Steinmacher        \and
        Marco A. Gerosa
}


\institute{Mairieli Wessel \at
           Delft University of Technology, The Netherlands \\
           The work was conducted while the author was affiliated to University of São Paulo\\
           \email{m.wessel@tudelft.nl} 
           \and
           Alexander Serebrenik \at
           Eindhoven University of Technology, The Netherlands \\
           \email{a.serebrenik@tue.nl}
           \and
           Igor Wiese, Igor Steinmacher \at
           Universidade Tecnologica Federal do Parana, Brazil \\
           \email{\{igor,igorfs\}@utfpr.edu.br}
           \and 
           Marco A. Gerosa \at
           Northern Arizona University \\
           \email{marco.gerosa@nau.edu}
}

\date{Received: date / Accepted: date}

\maketitle

\begin{abstract}
Software bots have been facilitating several development activities in Open Source Software (OSS) projects, including code review. However, these bots may bring unexpected impacts to group dynamics, as frequently occurs with new technology adoption. Understanding and anticipating such effects is important for planning and management. To analyze these effects, we investigate how several activity indicators change after the adoption of a code review bot. We employed a regression discontinuity design on 1,194 software projects from GitHub. We also interviewed 12 practitioners, including open-source maintainers and contributors. Our results indicate that the adoption of code review bots increases the number of monthly merged pull requests, decreases monthly non-merged pull requests, and decreases communication among developers. From the developers' perspective, these effects are explained by the transparency and confidence the bot comments introduce, in addition to the changes in the discussion focused on pull requests. Practitioners and maintainers may leverage our results to understand, or even predict, bot effects on their projects.
\end{abstract}

\section{Introduction}

Open Source Software (OSS) projects frequently employ code review in the development process~\cite{baysal2016investigating}, as it is a well-known practice for software quality assurance~\cite{ebert2019confusion}. In the pull-based development model, project maintainers carefully inspect code changes and engage in discussion with contributors to understand and improve the modifications before integrating them into the codebase~\cite{mcintosh2014impact}. The time maintainers spend reviewing pull requests is non-negligible and can affect, for example, the volume of new contributions~\cite{YuWaitforit} and the onboarding of newcomers~\cite{steinmacher2013newcomers}.

Software bots play a prominent role in the code review process~\cite{Wessel2018}. These automation tools serve as an interface between users and other tools~\cite{Storey2016} and reduce the workload of maintainers and contributors. Accomplishing tasks that were previously performed solely by human developers, and interacting in the same communication channels as their human counterparts, bots have become new voices in the code review conversation~\cite{Monperrus2019}. Throughout comments on pull requests, code review bots guide contributors to provide necessary information before maintainers triage the pull requests~\cite{Wessel2018}.

Notoriously, though, the adoption of new technology can bring consequences that counter the expectations of the technology designers and adopters~\cite{healy2012unanticipated}. Many systems intended to serve the user ultimately add new burdens. Developers who \emph{a priori} expect technological developments to produce significant performance improvements can be caught off-guard by \emph{a posteriori} unanticipated operational complexities~\cite{woods2001unexpected}. According to \citet{mulder2013impact}, many effects are not directly caused by the new technology itself, but by the changes in human behavior that it provokes. Therefore, it is important to assess and discuss the effects of a new technology on group dynamics; yet, this is often neglected when it comes to software bots~\cite{Storey2016,Paikari.vanDerHoek:2018}.

In the code review process, bots may affect existing project activities in several ways. For example, bots can provide poor feedback~\cite{Wessel2018, Wessel2020}, as illustrated by a developer: ``\textit{the comments of @$<$bot-name$>$ should contain more description on how to read the information contained and what one actually [understand] from it. For a newcomer its not obvious at all}.''\footnote{https://twitter.com/markusstaab/status/1048503185361555457} In turn, this poor feedback may lead to contributor drop-out---indeed, poor feedback on pull requests is known to discourage further contributions~\cite{Steinmacher.Pinto.ea_2018, balali2018newcomers}.

In this paper, we aim to understand how the dynamics of GitHub project pull requests change following the adoption of code review bots. To understand what happens after the adoption of a bot, we used a mixed-methods approach~\cite{easterbrook2008selecting} with a sequential explanatory strategy~\cite{creswell2003mixed}, combining data analysis of GitHub data with semi-structured interviews conducted with open-source developers. We used a \textit{Regression Discontinuity Design} (RDD)~\cite{thistlethwaite1960regression} to model the effects of code review bot adoption across 1,194 OSS projects hosted on GitHub. We used RDD since it can assess how much an intervention changed an outcome of interest, immediately and over time, and also evaluate whether the change could be attributed to other factors rather than the intervention. Afterward, to further shed light on our results, we conducted semi-structured interviews with practitioners, including open-source project maintainers and contributors experienced with code review bots.

We found that, after code review bot adoption, more pull requests are merged into the codebase, and communication decreases between contributors and maintainers. Considering non-merged pull requests, after bot adoption projects have fewer monthly non-merged pull requests, and faster pull request rejections. From the practitioners' perspective, the bot comments make it easier to understand the state and quality of the contributions and increase maintainers' confidence in merging pull requests. According to them, contributors are likely to make changes in the code without interacting with other maintainers, which also helps to change the focus of developers' discussions.

The main contributions of this paper are:

\begin{enumerate}
    \item The identification of changes in project activity indicators after the adoption of a code review bot.
    \item The elucidation of how the introduction of a bot can impact OSS projects.
    \item Open-source developers' perspective on the impacts of code reviews bots.
\end{enumerate}

These contributions aim to help practitioners and maintainers understand bots' effects on a project, especially to avoid the ones that they consider undesirable. Additionally, our findings may guide developers to consider the implications of new bots as they design them.

This paper extends our ICSME 2020 paper entitled ``Effects of Adopting Code Review Bots on Pull Requests to OSS Projects'' (\citet{wessel2020effects}). In this extended version, we further investigate the reasons for the change incurred by code review bot adoption, considering the practical perspective of open-source developers. To do so, we adjusted the methodology, results, and discussion, including a new research question (i.e., RQ2), which is based on the qualitative analysis of interviews with 12 open-source developers. We also present a more extensive related work section, where we discuss empirical works that use Regression Discontinuity Design to model the effects of a variety of interventions on development activities.

\section{Exploratory Case Study}
\label{sec:casestudy}

\begin{figure}[!htbp]
\scriptsize
\centerline{\includegraphics[scale=0.75]{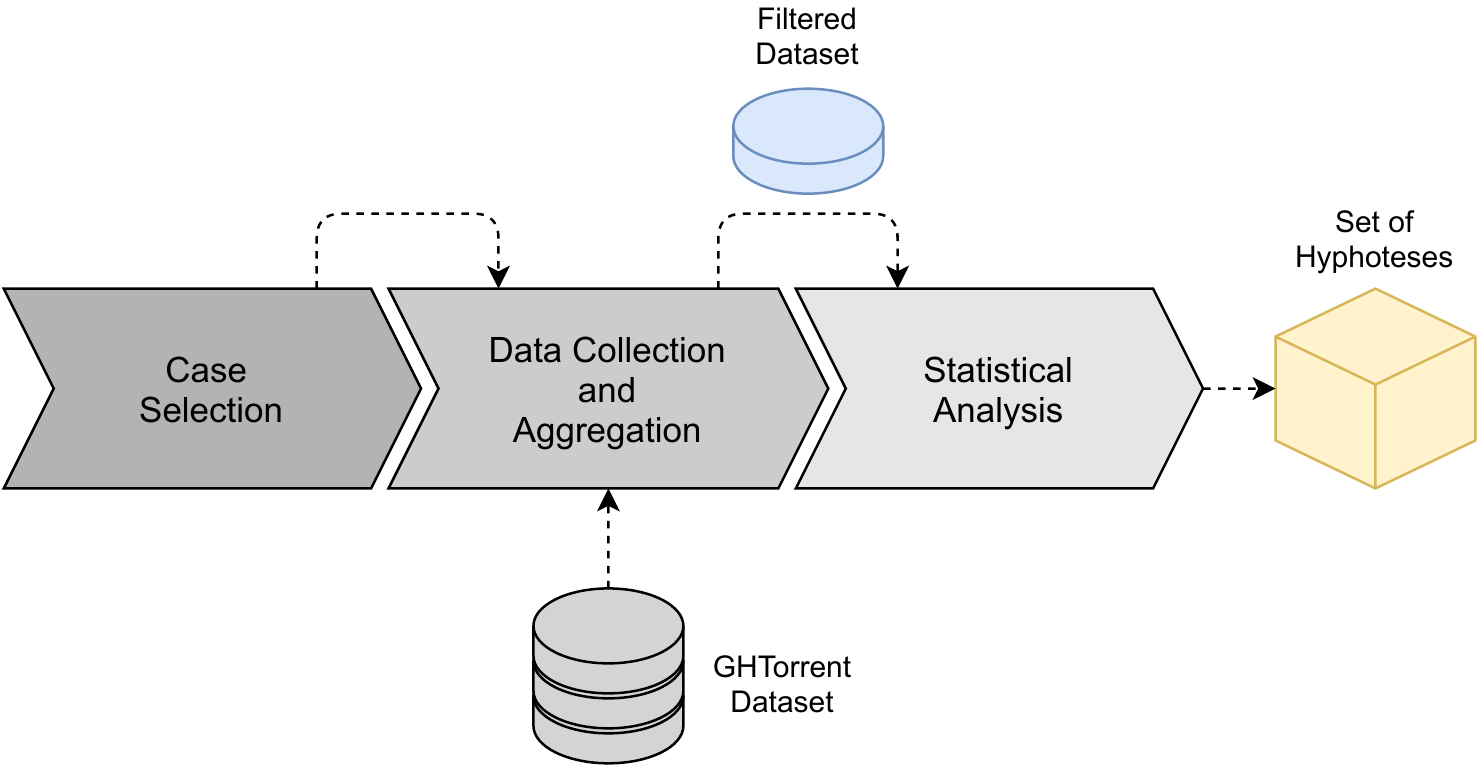}}
\caption{Case Study Design Overview.}
\label{fig:overview-casestudy}
\end{figure}

As little is known about the effects of code review bots' adoption in the dynamics of pull requests, we conducted an exploratory case study~\cite{runeson2009guidelines,yin2003design} to formulate hypotheses to further investigate in our main study. Figure~\ref{fig:overview-casestudy} shows an overview of the research design employed in this exploratory case study.

\subsection{Code Review Bot on Pull Requests}

\begin{figure}[htb]
\scriptsize
\centerline{\includegraphics[scale=0.7]{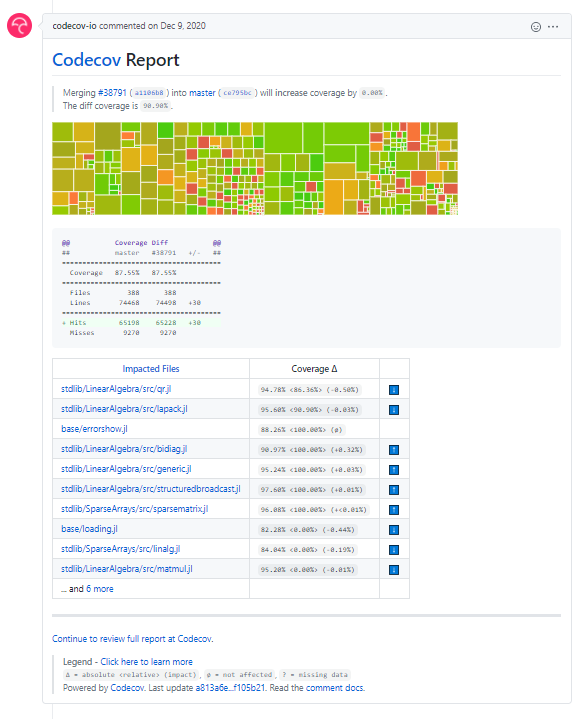}}
\caption{Codecov bot comment example.}
\label{fig:codecov}
\end{figure}

According to \citet{Wessel2018}, code review bots are software bots that analyze code style, test coverage, code quality, and smells. As an interface between human developers and other tools, code review bots generally report the feedback of a third-party service on the GitHub platform. Thus, these bots are designed to support contributors and maintainers after pull requests have been submitted, aiming to facilitate discussions and assist code reviews. One example of a code review bot is the Codecov bot.\footnote{https://github.com/marketplace/codecov} This bot reports the code coverage on every new pull request right after all tests have passed. As shown in Figure \ref{fig:codecov}, Codecov bot leaves highly detailed comments, including the percentage of increased or decreased coverage, and the impacted files.`

\subsection{Case Selection}

To carry out our exploratory case study, we selected two projects that we were aware of that used code review bots for at least a one year: the Julia programming language project\footnote{https://github.com/JuliaLang/julia} and CakePHP,\footnote{https://github.com/cakephp/cakephp} a web development framework for PHP. Both projects have popular and active repositories---Julia has more than $26.1k$ stars, $3.8k$ forks, $17k$ pull requests, and $46.4k$ commits; while CakePHP has more than $8.1k$ stars, $3.4k$ forks, $8.6k$ pull requests, $40.9k$ commits, and is used by $10k$ projects. Both projects adopt Codecov bot, which posted the first comments on pull requests to the Julia project in July 2016 and CakePHP in April 2016.

\subsection{Data Collection and Aggregation}

After selecting the projects, we analyzed data from one year before and one year after bot adoption, using the data available in the GHTorrent dataset~\cite{gousios2012ghtorrent}. During this time frame, the only bot adopted by Julia and CakePHP was the Codecov bot. Similar to previous work~\cite{zhao2017impact}, we exclude $30$ days around the bot's adoption to avoid the influence of instability caused during this period. Afterward, we aggregated individual pull request data into monthly periods, considering $12$ months before and after the bot's introduction. We choose the month time frame based on previous literature~\cite{zhao2017impact,kavaler2019tool,cassee2020silent}. All metrics were aggregated based on the month of the pull request being closed/merged.

We considered the same activity indicators used in the previous work by \citet{Wessel2018}:

\MyPara{Merged/non-merged pull requests:} the number of monthly contributions (pull requests) that have been merged, or closed but not merged into the project, computed over all closed pull requests in each time frame.

\MyPara{Comments on merged/non-merged pull requests:}the median number of monthly comments---excluding bot comments---computed over all merged and non-merged pull requests in each time frame.

\MyPara{Time-to-merge/time-to-close pull requests:} the median of monthly pull request latency (in hours), computed as the difference between the time when the pull request was closed and the time when it was opened. The median is computed using all merged and non-merged pull requests in each time frame.

\MyPara{Commits of merged/non-merged pull requests:} the median of monthly commits computed over all merged and non-merged pull requests in each time frame. 

For all activity indicators we use the median because their distribution is skewed.

\subsection{Statistical Analysis}

We ran statistical tests to compare the activity indicators distributions before and after the bot adoption. As the sample is small, and there is no critical mass of data points around the bot's introduction, we used the non-parametric Mann-Whitney-Wilcoxon test~\cite{wilks2011statistical}. In this context, the null hypothesis ($H_0$) is that the distributions pre- and post-adoption are the same, and the alternative hypothesis ($H_1$) is that these distributions differ. We also used Cliff's Delta~\cite{romano2006appropriate} to quantify the difference between these groups of observations beyond $p$-value interpretation. Moreover, we inspected the monthly distribution of each metric to search for indications of change.

As aforementioned, the case studies helped us to formulate hypotheses for the main study, which comprised more than one thousand projects. We formulated hypotheses whenever we observed changes in the indicators for at least one of the two projects we analyzed in the case study.

\subsection{Case Study Results}

In the following, we discuss the trends in project activities after bot adoption. We report the results considering the studied pull request activities: number of merged and non-merged pull requests, median of pull request comments, time-to-merge and time-to-close pull requests, and median of pull request commits.

\subsubsection{Trends in the number of Merged and Non-merged Pull Requests}

\begin{figure}[!htbp]
\scriptsize
\centerline{\includegraphics[scale=0.9]{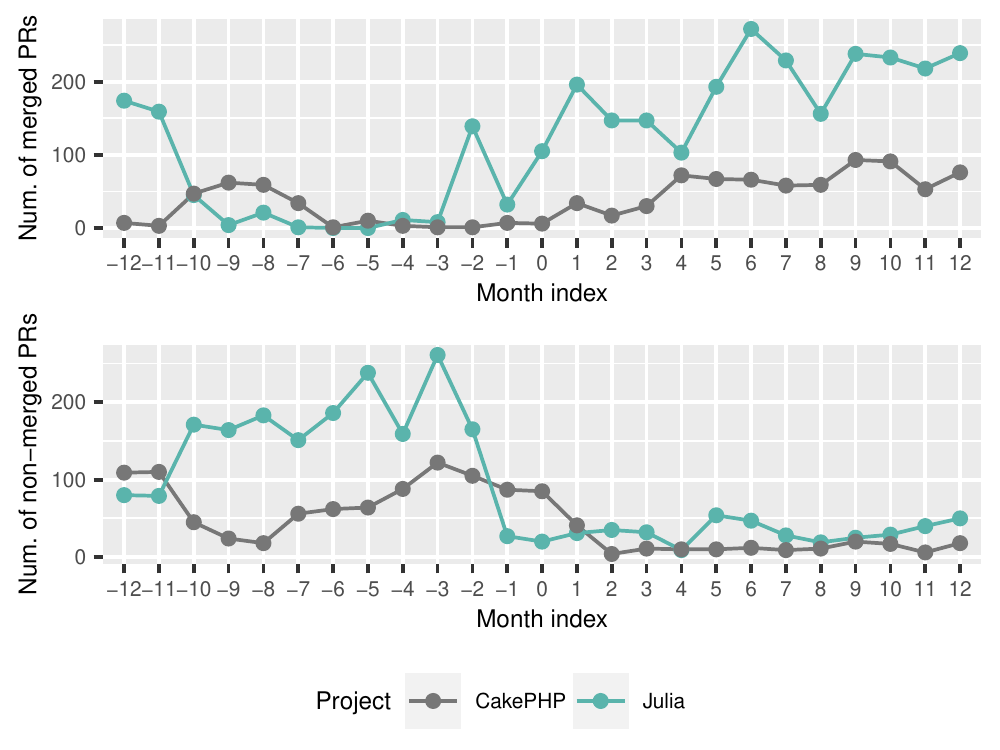}}
\caption{Monthly merged and non-merged pull requests.}
\label{fig:prs-julia}
\end{figure}

The number of merged pull requests \emph{increased} for both projects (Julia: $p$-value $0.0003$, $\delta=-0.87$; CakePHP: $p$-value $0.001$, $\delta=-0.76$), whereas the non-merged pull requests \emph{decreased} for both projects (Julia: $p$-value $0.00007$, $\delta=0.87$; CakePHP: $p$-value $0.00008$, $\delta =0.95$). Figure~\ref{fig:prs-julia} shows the monthly number of merged and non-merged pull requests, top and bottom respectively, before and after bot adoption for both projects. Based on these findings, we hypothesize that: 

\MyBox{\textbf{\textit{H$_{1.1}$ The number of monthly merged pull requests increases after the introduction of a code review bot.}}}

\MyBox{\textbf{\textit{H$_{1.2}$ The number of monthly non-merged pull requests decreases after the introduction of a code review bot.}}}

\subsubsection{Trends in the median of pull request comments}

\begin{figure}[!htbp]
\scriptsize
\centerline{\includegraphics[scale=0.9]{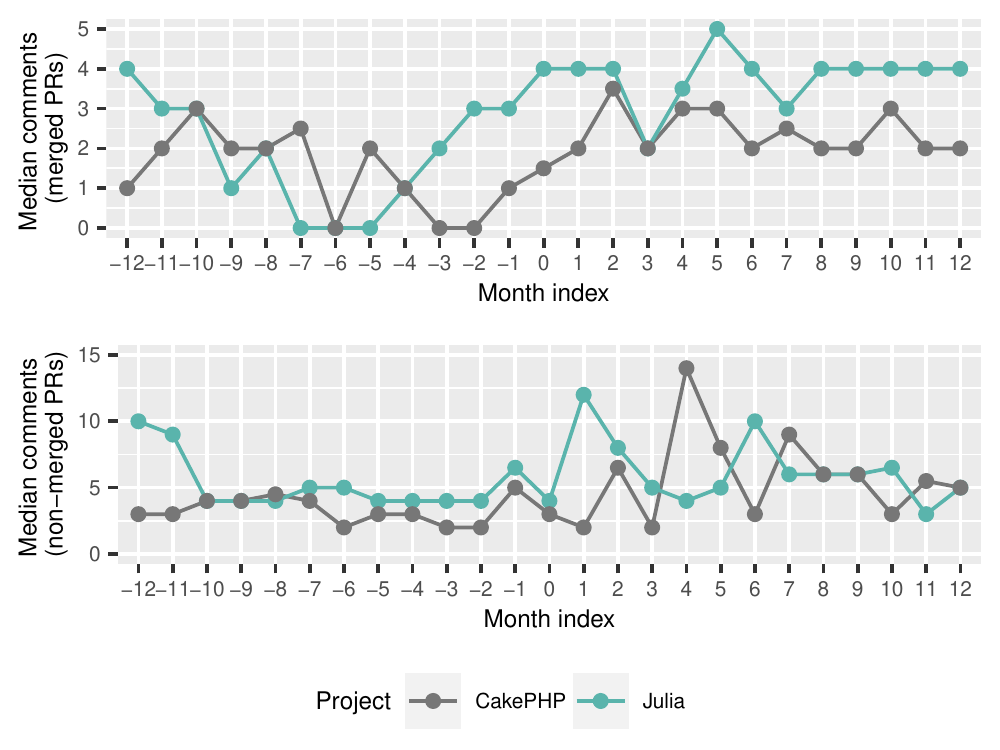}}
\caption{Monthly comments on merged and non-merged pull requests.}
\label{fig:comments-cakephp}
\end{figure}

Figure~\ref{fig:comments-cakephp} shows the monthly median of comments on merged and non-merged pull requests, respectively. CakePHP showed statistically significant differences between pre- and post-adoption distributions. The number of comments \emph{increased} for merged pull requests ($p$-value=$0.01$, $\delta =-0.56$) and also for non-merged ones ($p$-value=$0.03$, $\delta=-0.50$) with a large effect size. Thus, we hypothesize that:

\MyBox{\textbf{\textit{H$_{2.1}$ The adoption of code review bots is associated with an increase in the monthly number of comments for merged pull requests.}}}

\MyBox{\textbf{\textit{H$_{2.2}$ The number of monthly comments on non-merged pull requests increases after the adoption of a code review bot.}}}

\subsubsection{Trends in the time to close pull request comments}

\begin{figure}[!htbp]
\scriptsize
\centerline{\includegraphics[scale=0.9]{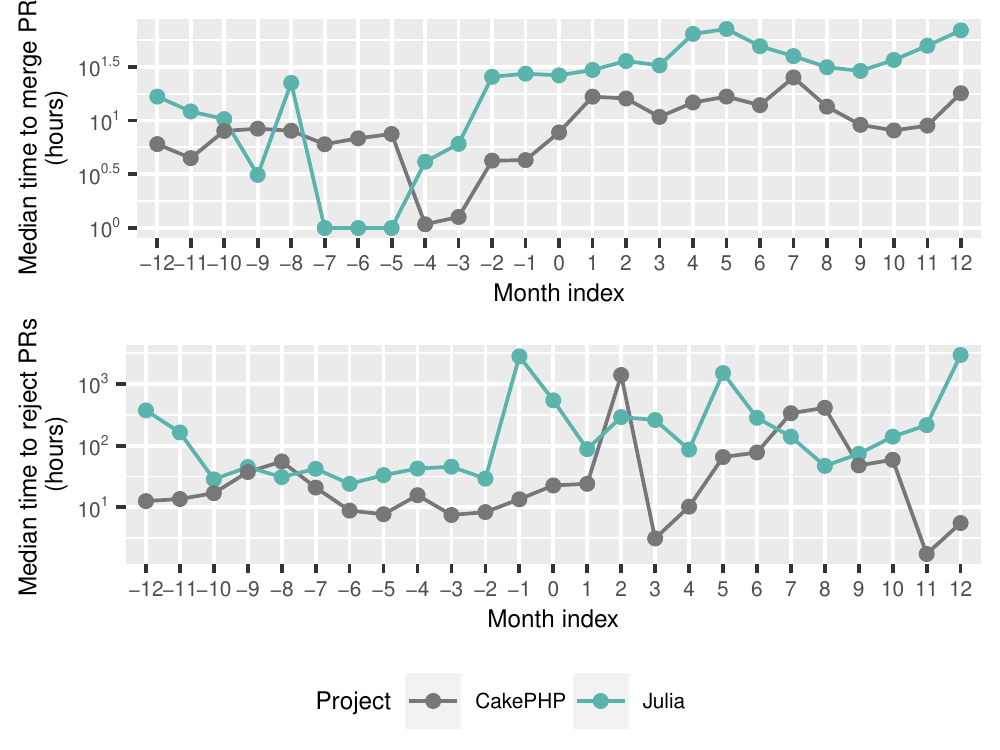}}
\caption{Monthly median time to merge and reject pull requests.}
\label{fig:time-julia}
\end{figure}

The median time to merge pull requests \emph{increased} for both projects (Julia: $p$-value $0.0003$, $\delta=-1.00$; CakePHP: $p$-value $0.000001$, $\delta=-0.98$). Considering non-merged pull requests, the difference between pre- and post-adoption is statistically significant only for Julia. For this project, the median time to close pull requests \emph{increased} ($p$-value $0.00007$) with a large effect size ($\delta=-0.65$). The distribution can be seen in Figure~\ref{fig:time-julia}. Therefore, we hypothesize that:

\MyBox{\textbf{\textit{H$_{3.1}$ There is an increase in the monthly time to merge pull requests after the introduction of code review bots.}}}

\MyBox{\textbf{\textit{H$_{3.2}$ There is an increase in the monthly time to reject pull requests after the adoption of a code review bot.}}}

\subsubsection{Trends in the median of pull request commits}

\begin{figure}[!htbp]
\scriptsize
\centerline{\includegraphics[scale=0.9]{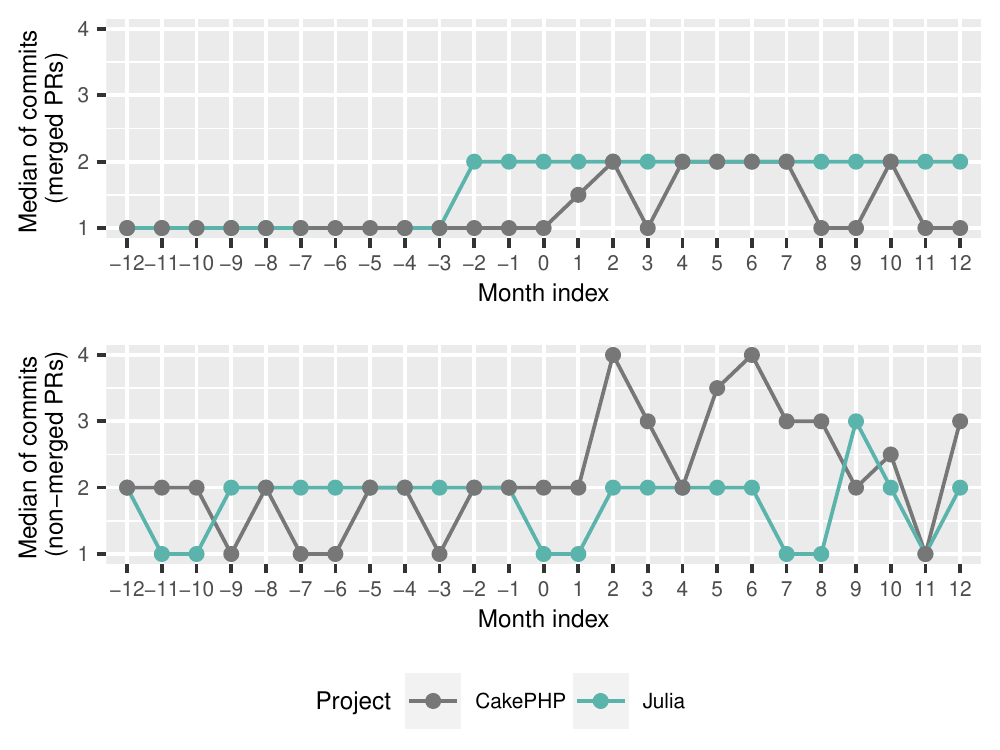}}
\caption{Monthly commits on merged and non-merged pull requests.}
\label{fig:commits-cakephp}
\end{figure}

Investigating the number of pull request commits per month (see Figure~\ref{fig:commits-cakephp}), we note that the medians before the adoption are quite stable, especially for merged pull requests. 
In comparison, after adoption we observe more variance.
The difference is statistically significant only for CakePHP, for which the number of pull request commits increased for merged pull requests ($p$-value=$0.002$, $\delta=-0.58$) and for non-merged pull requests ($p$-value=$0.002$, $\delta=-0.69$) with a large effect size. Based on this, we posit:

\MyBox{\textbf{\textit{H$_{4.1}$ There is an increase in the monthly number of commits for merged pull requests after code review bot adoption.}}}

\MyBox{\textbf{\textit{H$_{4.2}$ There is an increase in the monthly number of commits for non-merged pull requests after code review bot adoption.}}}

\MyBox{\textbf{Summary of the Case Study.} Unlike \citet{Wessel2018}, we observe statistically significant differences for all four activity indicators we investigated in at least one of the two projects. Based on these observations, we formulated hypotheses to be further investigated in our main study, comprising a large number of projects, and employed the regression discontinuity design.}

\section{Main Study Design}
\label{sec:method}
In this section, we describe our research questions (Section~\ref{sec-research-questions}), the statistical approach and data collection procedures (Section~\ref{sec-statistical-modeling}), and the qualitative approach (Section~\ref{sec-qualitative-approach}).

\subsection{Research Questions}
\label{sec-research-questions}

The main goal of this study is to investigate how and for what reasons, if any, the adoption of code review bots affects the dynamics of GitHub project pull requests. To achieve this goal, we investigated the following research questions:

\textbf{RQ1.} \textit{How do pull request activities change after a code review bot is adopted in a project?}

Extending the work of \citet{Wessel2018}, we investigate changes in project activity indicators, such as the number of pull requests merged and non-merged, number of comments, the time to close pull requests, and the number of commits per pull request. Using time series analysis, we account for how the bot adoption has impacted these project activity indicators over time. We also go one step further, exploring a large sample of open-source projects and focusing on understanding the effects of a specific bot category.

\textbf{RQ2.} \textit{How could the change in pull request activities be explained?}

Besides understanding the change incurred by bot adoption, we explore why it happens. To do so, we interviewed a set of open-source developers who actually have been using these bots.

\begin{figure}[!htbp]
\scriptsize
\centerline{\includegraphics[scale=0.75]{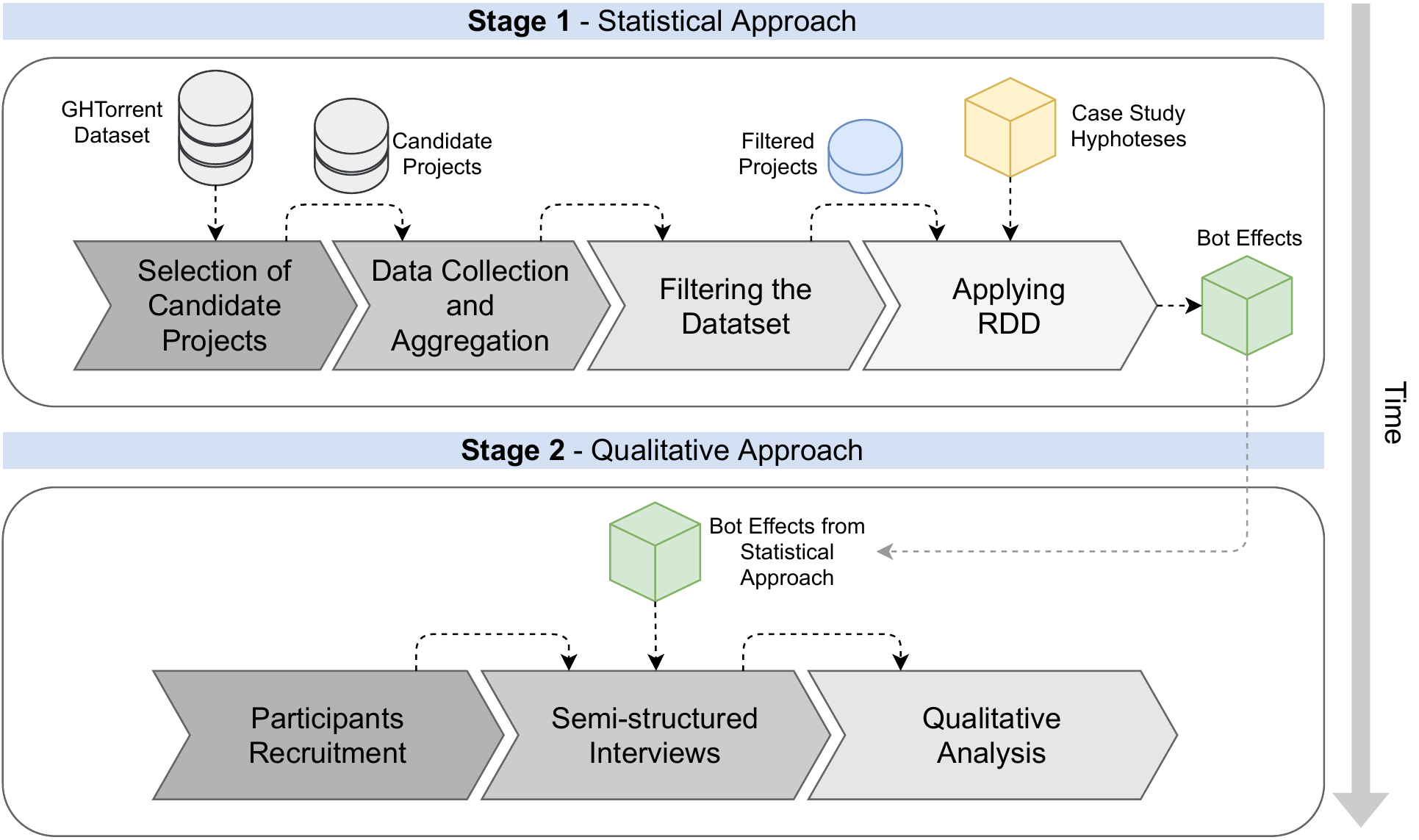}}
\caption{Main Research Design Overview.}
\label{fig:overview-main}
\end{figure}

Figure \ref{fig:overview-main} illustrates an overview of the steps taken to address the research questions. Next, we explain each step in order to justify the study design decisions.

\subsection{Stage 1---Statistical Approach}
\label{sec-statistical-modeling}

Considering the hypotheses formulated in the case study, in our main study we employed time series analysis to account for the longitudinal effects of bot adoption. We employed Regression Discontinuity Design (RDD)~\cite{thistlethwaite1960regression,imbens2008regression}, which has been applied in the context of software engineering in the past~\cite{zhao2017impact,cassee2020silent}. RDD is a technique used to model the extent of a discontinuity at the moment of intervention and long after the intervention. The technique is based on the assumption that if the intervention does not affect the outcome, there would be no discontinuity, and the outcome would be continuous over time~\cite{quasiexperimentation}. 
The statistical model behind RDD is
\begin{equation*}
\begin{split}
y_{i} =&\: \alpha + \beta\cdot \mbox{\textit{time}}_{i} + \gamma\cdot \mbox{\textit{intervention}}_{i} \: + \\& \delta\cdot \mbox{\textit{time\_after\_intervention}}_{i} \: + \eta\cdot \mbox{\textit{controls}}_{i} + \varepsilon_{i}
\end{split}
\end{equation*}
where $i$ indicates the observations for a given project.
To model the passage of time as well as the bot introduction, we include three additional variables: \textit{time}, \textit{time after intervention}, and \textit{intervention}. The \textit{time} variable is measured as months at the time $j$ from the start to the end of our observation period for each project ($24$ months). The \textit{intervention} variable is a binary value used to indicate whether the time $j$ occurs before ($\mbox{\textit{intervention}}=0$) or after ($\mbox{\textit{intervention}}=1$) adoption event. The \textit{time\_after\_intervention} variable counts the number of months at time $j$ since the bot adoption, and the variable is set up to $0$ before adoption.

The $\mbox{\textit{controls}}_{i}$ variables enable the analysis of bot adoption effects, rather than confounding the effects that influence the dependent variables. For observations before the intervention, holding controls constant, the resulting regression line has a slope of $\beta$, and after the intervention it has an slop of $\beta+\delta$. Further, the size of the intervention effect is measured as the difference equal to $\gamma$ between the two regression values of $y_{i}$ at the moment of the intervention.

Considering that we are interested in the effects of code review bots on the monthly trend of the number of pull requests, number of comments, time-to-close pull requests, and number of commits over a pull request, and all these for both merged and non-merged pull requests, we fitted eight models ($2$ cases x $4$ variables). To balance false-positives and false-negatives, we report the corrected p-values after applying multiple corrections using the method of Benjamini and Hochberg~\cite{benjamini1995controlling}. We implemented the RDD models as a mixed-effects linear regression using the R package \textit{lmerTest}~\cite{kuznetsova2017lmertest}.

To capture project-to-project and language-to-language variability, we modeled \textit{project name} and \textit{programming language} as random effects~\cite{galecki2013linear}. By modeling these features as random effects, we can account for and explain different behaviors observed across projects or programming languages~\cite{zhao2017impact}. We evaluate the model fit using \textit{marginal} $(R^2_m)$ and \textit{conditional} $(R^2_c)$ scores, as described by Nakagawa and Schielzeth~\cite{nakagawa2013general}. The $R^2_m$ can be interpreted as the variance explained by the fixed effects alone, and $R^2_c$ as the variance explained by the fixed and random effects together.

In mixed-effects regression, the variables used to model the intervention along with the other fixed effects are aggregated across all projects, resulting in coefficients useful for interpretation. The interpretation of these regression coefficients supports the discussion of the intervention and its effects, if any. Thus, we report the significant coefficients ($p < 0.05$) in the regression as well as their variance, obtained using ANOVA. In addition, we \textit{log} transform the fixed effects and dependent variables that have high variance~\cite{sheather2009modern}. We also account for multicollinearity, excluding any fixed effects for which the variance inflation factor (VIF) is higher than $5$~\cite{sheather2009modern}.

\subsubsection{Selection of Candidate Projects}
To identify open-source software projects hosted on GitHub that at some point had adopted a code review bot, we queried the GHTorrent dataset~\cite{gousios2012ghtorrent} and filtered projects in which at least one pull request comment was made by one of the code review bots identified by \citet{Wessel2018}. Following the method used by \citet{zhao2017impact} to assemble a time series, we considered only those projects that had been active for at least one year before and one year after the bot adoption. We found $4,767$ projects that adopted at least one of the four code review bots identified by \citet{Wessel2018} (ansibot, elasticmachine, codecov-io, coveralls). 
For each project, we collected data on all its merged and non-merged pull requests. By analyzing these projects we noticed that $220$ of them adopted both codecov-io and coveralls, while the other $4,547$ adopted only one of the code reviews bots (coveralls: 3,269; codecov-io: 1,270; elasticmachine: 5; ansibot: 3).

\subsubsection{Data Collection and Aggregation}
\label{ss:time:series}

Similar to the exploratory case study (see Section~\ref{sec:casestudy}), we aggregated the project data in monthly time frames and collected the four variables we expected to be influenced by the introduction of the bot: number of merged and non-merged pull requests, median number of comments, median time-to-close pull requests, and median number of commits. All these variables were computed over pull requests that have been merged and non-merged in a time frame.

We also collected six control variables, using the GHTorrent dataset~\cite{gousios2012ghtorrent}:

\MyPara{Project name:}the name of the project, used to identify the project on GitHub. We accounted for the fact that different projects can lead to different contribution patterns. We used the project name as a random effect.

\MyPara{Programming language:}the primary project programming language as automatically determined and provided by GitHub. We considered that projects with different programming languages can lead to different activities and contribution patterns~\cite{zhao2017impact,cassee2020silent}. We used programming language as a random effect.

\MyPara{Time since the first pull request:}in months, computed since the earliest recorded pull request in the entire project history. We use this to capture the difference in adopting the bot earlier or later in the project life cycle, after the projects started to use pull requests~\cite{zhao2017impact,cassee2020silent}.

\MyPara{Total number of pull request authors:}as a proxy for the size of the project community, we counted how many contributors submitted pull requests to the project.

\MyPara{Total number of commits:} as a proxy for the activity level of a project, we computed the total number of commits since the earliest recorded commit in the entire project history.

\MyPara{Number of pull requests opened:}the number of contributions (pull requests) received per month by the project. We expected that projects with a high number of contributions also observe a high number of comments, latency, commits, and merged and non-merged contributions.

\begin{table*}[!htbp]
\centering
\caption{An overview of the studied bots}
\label{tab:bots}
\begin{tabular}{llr}
\toprule
\textbf{Bot name} & \textbf{GitHub user} & \textbf{\# of projects} \\
\midrule
Ansible's issue bot
 & ansibot & $1$ \\ \hline
Elastic Machine & elasticmachine & $3$ \\ \hline
Codecov & codecov-io & $460$ \\ \hline
Coveralls & coveralls & $730$ \\
\midrule
\multicolumn{3}{r}{Total of $1,194$ under study}\\
\bottomrule
\end{tabular}
\end{table*}

\subsubsection{Filtering the final dataset}
After excluding the period of instability (30 days around the adoption), we inspected the dataset and found $223$ projects with no comments authored by any of the studied bots. We manually checked $30\%$ of these cases and concluded that some projects only added the bot for a testing period and then disabled it. We removed these $223$ projects from our dataset.

We also checked the activity level of the candidate projects, since many projects on GitHub are inactive~\cite{gousios2014exploratory}.
We excluded from our dataset projects without at least a six month period of consistent pull request activity during the one-year period before and after bot adoption. After applying this filter, a set of 1,740 GitHub software projects remained. To ensure that we observed the effects of each bot separately, we also excluded from our dataset 78 projects that adopted more than one of the studied bots and 196 projects that used non-code review bots. In addition, we checked the activity level of the bots on the candidate projects to remove projects that disabled the bot during the analyzed period. We then excluded 272 projects that had not received any comments during the previous four months. After applying all filters, 1,194 GitHub software projects remained. Table~\ref{tab:bots} shows the number of projects per bot. All of these four bots perform similar tasks on pull requests---providing comments on pull requests about code coverage.

\subsection{Stage 2---Qualitative approach}
\label{sec-qualitative-approach}

As aforementioned, we also applied a qualitative approach aimed to understand the effects evidenced by the statistical approach from the practitioners' perspective. In the following, we describe the participants recruitment, semi-structured interview procedures, and the qualitative analysis.

\subsubsection{Participants recruitment}

In this study, we employed several strategies to recruit participants. First, we advertised the interview on social media platforms frequently used by developers~\cite{singer2014software, storey2010impact, aniche2018modern}, including Twitter, Facebook, and Reddit. We also manually searched the projects that were part of the statistical analysis for pull requests explicitly installing or (re)configuring the analyzed bots. We added a comment on some of these pull requests to invite the pull request author to the interview. We also sent emails to personal contacts who we knew had experience with these bots. In addition, we asked participants to refer us to other qualified participants.

We continued recruiting participants till we came to an agreement that the last three interviews had not provided any new findings. According to Strauss and Corbin~\cite{strauss1997grounded}, sampling can be discontinued once the data collection no longer unveils new information. Additionally, the size of our participant set is in line with the anthropology literature, which mentions that a set of 10-20 knowledgeable people is sufficient to uncover and understand the core categories in any study of lived experience~\cite{bernard2017research}.

\subsubsection{Participants Demographics}

In total, we interviewed 12 open-source developers experienced with code review bots---identified here as P1--P12. Out of these twelve participants, one is an open-source maintainer, two are contributors, and the other nine are both maintainers and contributors. In addition, participants are geographically distributed across Europe ($\simeq$50\%), North America ($\simeq$25\%), and South America ($\simeq$25\%). Snowballing was the origin of five of our participants. Personal contacts was the origin of four of our participants. The advertisements on social media were the origin of the other three interviews. Table~\ref{tab:participants} presents the demographic information of the interviewees.

\begin{table}[!htbp]
\centering
\caption{Demographics of interviewees}
\label{tab:participants}
\begin{threeparttable}
\scriptsize
\begin{tabular}{cccccc}
\toprule
\textbf{Participant} & \textbf{OSS Experience} & \multicolumn{2}{c}{\textbf{Experienced with bots as}}  & \textbf{Location} & \textbf{Gender}\\
\textbf{ID} & \textbf{(years)} &  \textbf{Maintainer} & \textbf{Contributor} & &\\
\midrule
P1 & 4-5 & \ding{51} &  & North America & Man \\ \hline
P2 & Over 10 & \ding{51} & \ding{51} & North America & Man \\ \hline
P3 & 4-5 & \ding{51} & \ding{51} & Europe & Man \\ \hline
P4 & 3 & \ding{51} & \ding{51} & Europe & Man \\ \hline
P5 & 4-5 & \ding{51} & \ding{51} & Europe & Woman \\ \hline
P6 & Over 10 & \ding{51} & \ding{51} & North America & Man \\ \hline
P7 & 5-10 & \ding{51} & \ding{51} & Europe & Man \\ \hline
P8 & 4-5 & \ding{51} & \ding{51} & Europe & Man \\ \hline
P9 & 1 &  & \ding{51} & Europe & Man \\ \hline
P10 & 4-5 & \ding{51} & \ding{51} & South America & Man \\ \hline
P11 & 4-5 & \ding{51} & & South America & Man \\ \hline
P12 & Over 10 &  & \ding{51} & South America & Man \\
\bottomrule 
\end{tabular}
\end{threeparttable}
\end{table}

\subsubsection{Semi-structured interviews}

We conducted \emph{semi-structured} interviews, comprising open- and closed-ended questions designed to elicit foreseen and unexpected information and enable interviewers to explore interesting topics that emerged during the interview~\cite{hove2005experiences}. Before each interview, we shared a consent form with the participants asking for their agreement. By participants' requests, one interview (P11) was conducted via email. The other eleven interviews were conducted via video calls. The participants received a 25-dollar gift card as a token of appreciation for their time.

We started the interviews with a short explanation of the research objectives and guidelines, followed by demographic questions to capture the familiarity of the interviewees with open-source development and code review bots. We then described to the interviewee the study we conducted and the main findings from the statistical approach and asked the developers to conjecture about the reasons for the effects we observed: 

\begin{itemize}
    \item[Q1.] After adopting a code review bot there are more merged pull requests, less communication between developers, fewer rejected pull requests, and faster rejections. We are intrigued about these effects and would like to hear thoughts from developers who actually use these bots. Could you conjecture the reasons why this happens?
\end{itemize}

We follow-up this question with more specific questions when participants have not mentioned reasons for any of the four observed effects. Afterwards, we asked two additional questions: 

\begin{itemize}
    \item[Q2.] Have you observed these effects in your own project?
    \item[Q3.] What other effects did you observe in your project and attribute to the introduction of the code review bot?
\end{itemize}

The detailed interview script is publicly available\footnotemark. Each interview was conducted remotely by the first author of this paper and lasted, on average, 35 minutes. 

\subsubsection{Qualitative analysis of interviews}

Each interview recording was transcribed by the first author of this paper. We then analyzed the interview transcripts by applying open and axial coding procedures~\cite{strauss1998basics,stol2016grounded} throughout multiple rounds of analysis. We started by applying open coding, whereby we identified the reasons for bots' effects. To do so, the first author of this paper conducted a preliminary analysis, identifying the main codes. Then, the first author discussed with fourth and fifth authors the coding in weekly hands-on meetings. These discussions aimed to increase the reliability of the results and mitigate bias~\cite{Strauss.Corbin_1998, patton2014qualitative}. Afterwards, the first author further analyzed and revised the interviews to identify relationships between concepts that emerged from the open coding analysis (axial coding). During this process, we employed a constant comparison method~\cite{glaser2017discovery}, wherein we continuously compared the results from one interview with those obtained from the previous ones. The axial coding resulted on grouping the participants' answers into five categories.

For confidentiality reasons, we do not share the interview transcripts. However, we made our complete code book publicly available. The code book includes the all code names, descriptions, and examples of quotes.

\footnotetext{https://doi.org/10.5281/zenodo.4618498}

\section{Main Study Results}
\label{sec:results}

In the following, we report the results of our study by research question.

\subsection{Effects of Code Review Bot Adoption (RQ1)}

In this section, we discuss the effects of code review bot adoption on project activities along four dimensions: (i) accepted and rejected pull requests, (ii) communication, (iii) pull request resolution efficiency, and (iv) modification effort.

\subsubsection{Effects in Merged and Non-merged Pull Requests}
\label{sub:merged_nonmerged}

We start by investigating the effects of bot adoption on the number of merged and non-merged pull requests. From the exploratory case study, we hypothesized that the use of code review bots is associated with an increase in the number of monthly merged pull requests and a decrease in the number of monthly non-merged pull requests. We fit two mixed-effect RDD models, as described in Section \ref{sec-statistical-modeling}. For these models, the \textit{number of merged/non-merged pull requests} per month is the dependent variable. Table~\ref{tab:resultspullrequest} summarizes the results of these two RDD models. In addition to the model coefficients, the table also shows the SS, with a variance explained for each variable. We also highlighted the time series predictors \textit{time}, \textit{time after intervention}, and \textit{intervention} in \textbf{bold}.

\begin{table*}[htbp]
\centering
\caption{The Effects of Code Review bots on PRs. The response is \textbf{log(number of merged/non-merged PRs)} per month.}
\label{tab:resultspullrequest}
\begin{threeparttable}
\begin{tabular}{lrrrrrr}
\toprule
 & \multicolumn{2}{c}{Merged Pull Requests} & & \multicolumn{2}{c}{Non-merged Pull Requests}\\
\cmidrule{2-3}\cmidrule{5-6}
 & Coefficients & SS & & Coefficients & SS \\
\cmidrule{2-3}\cmidrule{5-6}
Intercept & -0.262*** & & & -0.574*** &\\
\rowcolor{Gray}
TimeSinceFirstPullRequest & 0.00004** & 4.3 & & -0.0001*** & 2.4 \\
log(TotalPullRequestAuthors) & -0.094*** & 171.8 & & 0.086*** & 775.7 \\
\rowcolor{Gray}
log(TotalCommits) & 0.042*** & 484.0 & & 0.068*** & 428.6\\
log(OpenedPullRequests) & 0.494*** & 8227.1 & & 0.388*** & 4958.5 \\
\rowcolor{Gray}
log(PullRequestComments) & 0.433*** & 2954.3 & & 0.389*** & 2341.0 \\
log(PullRequestCommits) & 0.272*** & 721.0 & & 0.165*** & 255.5 \\
\rowcolor{Gray}
\textbf{time} & 0.004*** & 203.2 & & -0.004*** & 376.1 \\
\textbf{interventionTrue} & 0.095*** & 16.8 & & -0.163*** & 48.4 \\
\rowcolor{Gray}
\textbf{time\_after\_intervention} & 0.004** & 1.7 && -0.004** & 1.6 \\
\midrule
Marginal $R^2$ & & 0.68 & & & 0.67\\
Conditional $R^2$ & & 0.75 & & & 0.74\\
\bottomrule
\end{tabular}
\begin{tablenotes}
 \item *** $p < 0.001$, ** $p < 0.01$, * $p < 0.05$. SS stands for ``Sum of Squares''.
 \item Time series predictors in \textbf{bold}.
\end{tablenotes}
\end{threeparttable}
\end{table*}

Analyzing the model for merged pull requests, we found that the fixed-effects part fits the data well ($R^2_m=0.68$). However, considering $R^2_c=0.75$, variability also appears from project-to-project and language-to-language. Among the fixed effects, we observe that the number of monthly pull requests explains most of the variability in the model. As expected, this indicates that projects receiving more contributions tend to have more merged pull requests, with other variables held constant.

Furthermore, the statistical significance of the time series predictors indicates that the adoption of code review bots affected the trend in the number of merged pull requests. Observing the \textit{time} coefficient, we note an increasing trend before adoption. There is a statistically significant discontinuity at adoption, since the coefficient for \textit{intervention} is statistically significant. Further, there is a positive trend after adoption (see \textit{time after intervention}) and the sum of the coefficients for \textit{time} and \textit{time after intervention} is positive; thus, indicating that the number of merged pull requests increased even faster after bot adoption.

Similar to the previous model, the fixed-effect part of the non-merged pull requests model fits the data well ($R^2_m=0.67$), even though a considerable amount of variability is explained by random effects ($R^2_c=0.74$). We note similar results on fixed effects: projects receiving more contributions tend to have more non-merged pull requests. All the three time-series predictors for this model are statistically significant, showing a measurable effect of the code review bot's adoption on the time to review and accept a pull request. The \textit{time} coefficient shows a decreasing trend before adoption, \textit{intervention} coefficient reports a statistically significant discontinuity at the adoption time, and there is a slight acceleration after adoption in the decreasing time trend seen before adoption observed since the sum of the coefficients for \textit{time} and \textit{time after intervention} is negative.

Therefore, based on models for merged and non-merged pull requests, we confirm both \textbf{H$_{1.1}$} and \textbf{H$_{1.2}$}.

\MyBox{\textbf{Effects in Merged and Non-merged Pull Requests.} Overall, there are more monthly merged pull requests and fewer monthly non-merged pull requests after adopting a code review bot.}

\subsubsection{Effects on Developers' Communication}

In the exploratory case study, we hypothesized that bot adoption increases monthly human communication on pull requests for both merged and non-merged pull requests. To statistically investigate this, we fit one model to merged pull requests and another to non-merged ones. The \textit{median of pull request comments} per month is the dependent variable, while \textit{number of monthly pull requests}, \textit{median of time-to-close pull requests}, and \textit{median of pull request commits} are independent variables. Table~\ref{tab:resultscomments} shows the results of the fitted models.

\begin{table*}[htbp]
\centering
\caption{The Effect of Code Review bots on Pull Request Comments. The response is \textbf{log(median of comments)} per month.}
\label{tab:resultscomments}
\begin{threeparttable}
\begin{tabular}{lrrrrrr}
\toprule
 & \multicolumn{2}{c}{Merged Pull Requests} & & \multicolumn{2}{c}{Non-merged Pull Requests}\\
\cmidrule{2-3}\cmidrule{5-6}
 & Coefficients & SS & & Coefficients & SS \\
\cmidrule{2-3}\cmidrule{5-6}
Intercept & -0.096*** & & & -0.123*** &\\
\rowcolor{Gray}
TimeSinceFirstPullRequest & 0.00000 & 20.0 & & -0.00002* & 24.4 \\
log(TotalPullRequestAuthors) & 0.053*** & 163.6 & & 0.069*** & 621.1 \\
\rowcolor{Gray}
log(TotalCommits) & -0.014*** & 36.6 & & -0.009** & 106.0 \\
log(OpenedPullRequests) & 0.079*** & 1002.8 & & 0.072*** & 1362.9 \\
\rowcolor{Gray}
log(TimeToClosePullRequests) & 0.093*** & 3239.7 & & 0.101*** & 4615.5 \\
log(PullRequestCommits) & 0.093*** & 55.0 & & 0.123*** & 119.4 \\
\rowcolor{Gray}
\textbf{time} & -0.001 & 1.0 & & -0.001 & 7.2 \\
\textbf{interventionTrue} & 0.023** & 0.8 & & -0.025*** & 1.1 \\
\rowcolor{Gray}
\textbf{time\_after\_intervention} & -0.002* & 0.5 && 0.0001 & 0.0\\
\midrule
Marginal $R^2$ & & 0.50 & & & 0.66\\
Conditional $R^2$ & & 0.56 & & & 0.70\\
\bottomrule
\end{tabular}
\begin{tablenotes}
 \item *** $p < 0.001$, ** $p < 0.01$, * $p < 0.05$. SS stands for ``Sum of Squares''.
 \item Time series predictors in \textbf{bold}.
\end{tablenotes}
\end{threeparttable}
\end{table*}

Considering the model of comments on merged pull requests, we found that the model taking into account only fixed effects ($R^2_m=0.50$) fits the data well. However, there is also variability from the random effects ($R^2_c=0.56$). We observe that \textit{time-to-close pull requests explains the largest amount of variability in the model}, indicating that communication during the pull request review is strongly associated with the time to merge it. Regarding the bot effects, there is a discontinuity at adoption time, followed by a statistically significant decrease after the bot's introduction.

As above, the model of non-merged pull requests fits the data well ($R^2_m=0.66$) and there is also variability explained by the random variables ($R^2_c=0.70$). This model also suggests that communication during the pull request review is strongly associated with the time to reject the pull request. Table~\ref{tab:resultscomments} shows that the effect of bot adoption on non-merged pull requests differs from the effect on merged ones. The statistical significance of the \textit{intervention} coefficient indicates that the adoption of code review bots slightly affected communication; however, there is no bot effect in the long run.

Since our model for merged pull requests shows a decrease in the number of comments after bot adoption, we rejected \textbf{H$_{2.1}$}. Still, given that our model for non-merged pull requests could not observe any statistically significant bot effect as time passes, we cannot accept \textbf{H$_{2.2}$}. 

\MyBox{\textbf{Effects in Communication.} On average, there is less monthly communication on merged pull requests after adopting a code review bot. However, the monthly communication on non-merged pull requests does not change as time passes.}

\subsubsection{Effects in Pull Request Resolution Efficiency}

In the exploratory case study, we found that the monthly time to close pull requests increased after bot adoption. Next, we fitted two RDD models, for both merged and non-merged pull requests, where \textit{median of time to close pull requests} per month is the dependent variable. The results are shown in Table~\ref{tab:resultstime}.

\begin{table*}[htbp]
\centering
\caption{The Effect of Code Review bots on time-to-close PRs. The response is \textbf{log(median of time-to-close PRs)} per month.}
\label{tab:resultstime}
\begin{threeparttable}
\begin{tabular}{lrrrrrr}
\toprule
 & \multicolumn{2}{c}{Merged Pull Requests} & & \multicolumn{2}{c}{Non-merged Pull Requests}\\ 
\cmidrule{2-3}\cmidrule{5-6}
 & Coefficients & SS & & Coefficients & SS \\
\cmidrule{2-3}\cmidrule{5-6}
Intercept & 0.377** & & & 0.221 &\\
\rowcolor{Gray}
TimeSinceFirstPullRequest & 0.0002** & 452 & & 0.00001 & 891 \\
log(TotalPullRequestAuthors) & 0.208*** & 2186 & & 0.166*** & 21320 \\
\rowcolor{Gray}
log(TotalCommits) & -0.145*** & 824 & & -0.057** & 4770 \\
log(OpenedPullRequests) & 0.120*** & 34444 & & 0.240*** & 50376 \\
\rowcolor{Gray}
log(PullRequestComments) & 2.472*** & 117571 & & 3.326*** & 176312 \\
log(PullRequestCommits) & 2.275*** & 47117 & & 1.721*** & 26733 \\
\rowcolor{Gray}
\textbf{time} & 0.027*** & 3007 & & 0.012** & 56 \\
\textbf{interventionTrue} & 0.256*** & 128 & & -0.056 & 9 \\
\rowcolor{Gray}
\textbf{time\_after\_intervention} & 0.009 & 6 && -0.028*** & 66 \\
\midrule
Marginal $R^2$ & & 0.61 & & & 0.69\\
Conditional $R^2$ & & 0.67 & & & 0.72\\
\bottomrule
\end{tabular}
\begin{tablenotes}
 \item *** $p < 0.001$, ** $p < 0.01$, * $p < 0.05$. SS stands for ``Sum of Squares''.
 \item Time series predictors in \textbf{bold}.
\end{tablenotes}
\end{threeparttable}
\end{table*}

Analyzing the results of the effect of code review bots on the latency to merge pull requests, we found that combined fixed-and-random effects fit the data better than the fixed effects only ($R^2_c = 0.67$ vs $R^2_m = 0.61$). Although several variables affect the trends of pull request latency, communication during the pull requests is responsible for most of the variability in the data. This indicates the expected results: the more effort contributors expend discussing the contribution, the more time the contribution takes to merge. The number of commits also explains the amount of data variability, since a project with many changes needs more time to review and merge them. Moreover, we observe an increasing trend before adoption, followed by a statistically significant discontinuity at adoption. After adoption, however, there is no bot effect on the time to merge pull requests since the \textit{time\_after\_intervention} coefficient is not statistically significant.

Turning to the model of non-merged pull requests, we note that it fits the data well ($R^2_m=0.69$), and there is also a variability explained by the random effects ($R^2_c=0.72$). As above, communication during the pull requests is responsible for most of the variability encountered in the results. In this model, the number of received contributions is important to explain variability in the data---projects with many contributions need more time to review and reject them. The effect of bot adoption on the time spent to reject pull requests differs from the previous model. Regarding the time series predictors, the model did not detect any discontinuity at adoption time. However, the positive trend in the latency to reject pull requests before bot adoption is reversed toward a decrease after adoption.

Thus, since we could not observe statistically significant bot effects as time passes, we cannot confirm \textbf{H$_{3.1}$}. Further, as the model of non-merged pull requests shows a decrease in the monthly time to close pull requests, we reject \textbf{H$_{3.2}$}.

\MyBox{\textbf{Effects in PR Resolution Efficiency.} After adopting the code review bot, on average less time is required from maintainers to review and reject pull requests. However, the time required to review and accept a pull request does not change after code review bot adoption.}

\subsubsection{Effects in Commits}
\label{sec:effectsincommits}

Finally, we studied whether code review bot adoption affects the number of commits made before and during pull request review. Our hypothesis is that the monthly number of commits increases with the introduction of code review bots. Again, we fitted two models for merged and non-merged pull requests, where the \textit{median of pull request commits} per month is the dependent variable. The results are shown in Table~\ref{tab:resultscommits}. 

\begin{table*}[htbp]
\centering
\caption{The Effect of Code review bots on Pull Request commits. The response is \textbf{log(median of Pull Request commits)} per month.}
\label{tab:resultscommits}
\begin{threeparttable}
\begin{tabular}{lrrrrrr}
\toprule
 & \multicolumn{2}{c}{Merged Pull Requests} & & \multicolumn{2}{c}{Non-merged Pull Requests}\\
\cmidrule{2-3}\cmidrule{5-6}
 & Coefficients & SS & & Coefficients & SS \\
\cmidrule{2-3}\cmidrule{5-6}
Intercept & 0.358*** & & & 0.063 &\\
\rowcolor{Gray}
TimeSinceFirstPullRequest & 0.0001*** & 0.30 & & 0.00002 & 5.7 \\
log(TotalPullRequestAuthors) & -0.144*** & 0.02 & & -0.058*** & 202.2 \\
\rowcolor{Gray}
log(TotalCommits) & 0.017*** & 74.04 & & 0.028*** & 171.9 \\
log(OpenedPullRequests) & 0.163*** & 1513.60 & & 0.125*** & 1502.9 \\
\rowcolor{Gray}
log(PullRequestComments) & 0.520*** & 2375.74 & & 0.600*** & 3306.3 \\
\textbf{time} & 0.001 & 138.60 & & -0.003** & 8.7 \\
\rowcolor{Gray}
\textbf{interventionTrue} & 0.137*** & 33.57 & & 0.003 & 0.0 \\
\textbf{time\_after\_intervention} & 0.001 & 0.05 && 0.001 & 0.1 \\
\midrule
Marginal $R^2$ & & 0.34 & & & 0.42\\
Conditional $R^2$ & & 0.48 & & & 0.50\\
\bottomrule
\end{tabular}
\begin{tablenotes}
 \item *** $p < 0.001$, ** $p < 0.01$, * $p < 0.05$. SS stands for ``Sum of Squares''.
 \item Time series predictors in \textbf{bold}.
\end{tablenotes}
\end{threeparttable}
\end{table*}

Analyzing the model of commits on merged pull requests, we found that the combined fixed-and-random effects ($R^2_c=0.48$) fit the data better than the fixed effects ($R^2_m=0.34$), showing that most of the explained variability in the data is associated with project-to-project and language-to-language variability, rather than with the fixed effects. The statistical significance of the \textit{intervention} coefficient indicates that the adoption of code review bots affected the number of commits only at the moment of adoption. Additionally, from Table~\ref{tab:resultscommits}, we can also observe that the number of pull request comments per month explains most of the variability in the result. This result suggests that the more comments there are, the more commits there will be, as discussed above. 

Investigating the results of the non-merged pull request model, we found that the model fits the data well and that the random effects are again important in this regard. We also observe from Table~\ref{tab:resultscommits} that the adoption of a bot is not associated with the number of commits on non-merged pull requests, since \textit{intervention} and \textit{time\_after\_intervention} coefficients are not statistically significant. 

Based on models for merged and non-merged pull requests, we could not observe statistically significant effects of bot adoption. Therefore, we cannot confirm both \textbf{H$_{4.1}$} and \textbf{H$_{4.2}$}.

\MyBox{\textbf{Effects in Commits.} After adopting a code review bot, the monthly trend in the median of pull request commits does not change for both merged and non-merged pull requests.}

\subsection{Developers' Perspective on the Reasons for the Observed Effects (RQ2)}

As explained in Section \ref{sec-qualitative-approach}, we presented to open-source developers the main findings of our statistical approach: ``\textit{After adopting a code review bot there are more merged pull requests, less communication between developers, fewer rejected pull requests, and faster rejections}.'' We asked them to conjecture on the possible reasons for each of these results. 

We grouped the participants' answers into 5 categories, as can be seen in Table~\ref{tab:reasons}. We associate one of the effects with its correspondent reasons whenever participants explicitly mentioned this relationship. We also added a mark (\ding{51}) to highlight which effects are explained by each one of the reasons, according to the participants' responses.

\begin{table}[!htbp]
\centering
\caption{Main reasons for the findings from the RDD models.}
\label{tab:reasons}
\scriptsize
\scalefont{0.8}
\begin{tabular}{|l|r|c|c|c|c|}
\hline
\multirow{4}{*}{\textbf{Reason}} & \multirow{4}{*}{\textbf{\#}}& \multicolumn{4}{c|}{\textbf{Explains}} \\ \cline{3-6} 
  & & \textbf{More} & \multirow{2}{*}{\textbf{Fewer}} & \textbf{Fewer} & \multirow{2}{*}{\textbf{Faster}} \\
    & & \textbf{Merged} & \multirow{2}{*}{\textbf{Comments}} & \textbf{Rejected} & \multirow{2}{*}{\textbf{Rejections}} \\
  & & \textbf{PRs} & & \textbf{PRs} & \textbf{} \\\hline\hline
More visibility and transparency & \multirow{2}{*}{8} & \multirow{2}{*}{\ding{51}} & \multirow{2}{*}{\ding{51}} & \multirow{2}{*}{\ding{51}} & \multirow{2}{*}{\ding{51}}\\ 
of the contribution state  &  &  &  &  &  \\\hline
More confidence in the & \multirow{2}{*}{8} & \multirow{2}{*}{\ding{51}} & \multirow{2}{*}{\ding{51}} & & \multirow{2}{*}{\ding{51}} \\
process in place & & & & & \\ \hline
Bot feedback changes developers' & \multirow{2}{*}{8} & & \multirow{2}{*}{\ding{51}} &  & \\
discussion focus & & & & & \\\hline
Bot feedback pushes contributors & \multirow{2}{*}{5} & \multirow{2}{*}{\ding{51}} & & \multirow{2}{*}{\ding{51}} &  \\
to take an action & & & & & \\ \hline
Bot feedback perceived as noise & 2 &  & \ding{51} &  & \\ \hline
\end{tabular}
\end{table}

\MyPara{More visibility and transparency of the contribution state.} Most of the participants claimed that when a project has bots that provide detailed information on code quality metrics, especially in the sense of coverage metrics, both maintainers and contributors can more quickly gain a general idea of the quality of the contributions. As stated by P6: ``\textit{bots are able to raise visibility, both for the contributor and for the maintainer. They can make it more clear more quickly the state of that contribution}.'' More than obtaining clarity on the quality of the code, it is also easy for maintainers to verify whether the pull request contributors will improve their contribution toward achieving acceptance. Thus, they conjecture that all bot effects we found during the statistical analysis might be explained by this enhanced feedback given by the bot.

As soon as contributors submit their pull requests, the code review bot posts a detailed comment regarding the code coverage. In the P7 experience, the ``\textit{immediate feedback of the quality of [code coverage] on the pull request}'' is closely related to the increasing acceptance rate. If the pull request does not affect code coverage in a negative way, then maintainers are able to ``\textit{much more quickly judge whether or not it's a reasonable request}'' (P4). On the other side of the spectrum, if the pull requests fail the tests and decreases the coverage, then the maintainers ``\textit{will not bother with that pull request at all, and just reject it}'' (P4). Also maintainers ``\textit{are more inclined to directly reject the pull request}'' since it does not respect the rules imposed by the project. In some cases, maintainers expect that the contributor will take an action based on the bot comments, as explained by P6: ``\textit{if [contributors] are not following up and resolving the issue, it makes it more clear to the maintainer that it's not an acceptable contribution.}''

Participants also recognize that these bots are usually ``\textit{pretty good at explaining very precisely}'' (P2) and not merely stating that ``\textit{[maintainers] will not accept the pull request}'' (P2) without further explanation. For example, if the coverage decreased, the bot will post ``\textit{your pull request dropped the test coverage from 95 to 94\%. And these are the lines you edit that are not covered. So, please add tests to cover these specific lines.}'' (P2), which according to P2 is extremely useful for a contributor. According to P1, for example, the visibility of the bot comments helps maintainers to make sure contributors understand why the pull request has been rejected without the necessity of engaging in a long discussion: ``\textit{now the maintainer can just point at it and be like `you didn't pass the status check, because you didn't write tests.' It is more obvious}''.

\MyPara{More confidence in the process in place.} According to the participants, one of the reasons for more pull requests being merged after the code review bot introduction is that these bots act as quality gatekeepers. For example, P1 mentioned that \textit{``by having other metrics, like code coverage, to be able to say `Great! I know that at least a test has been written for that line of code', there is some sort of gatekeeping.''} Besides the effect of merging more pull requests, participants also mentioned another effect: ``\textit{accepting code contributions can be much, much faster}'' (P2). Basically, code review bots are used as a way to achieve ``\textit{automatic verification}'' (P7). According to P7, if the bots confirm that the change is correct, then ``\textit{the developer is more convinced that the change is useful and valid.}'' In the opposite way, if the bots shows that the change is incorrect, the pull request will be rejected faster, as it does not require ``\textit{human interaction to arrive at this conclusion}'' (P7), which implies less communication between developers. Furthermore, P4 also relates the confidence in the bot as one of the reasons for less communication between developers: \textit{the fact that there is less communication between the contributors and maintainers might be an effect that we can get a bit overdependent on bots, in the sense you trust them too much.}'' Therefore, since maintainers trust the bots' feedback, they ``\textit{ask fewer questions}'' (P2).

\MyPara{Bot feedback changes developers' discussion focus.} Participants recurrently mentioned that bot comments enabled them to focus on other high-priority discussions, which led to a decrease in the communication between the project maintainers and contributors on pull requests. To some extent this decrease occurs since \textit{it's not necessary anymore, because a lot of that [comments] are [already] handled automatically by the bots}'' (P4). In P3's experience, maintainers ``\textit{talk more for new developers, to text them usually things like `Add new test please' and then [maintainers] don't have to [make] that kind of comment[] anymore. That's why there's less communication}.'' 

Moreover, when receiving non-human feedback, contributors are less likely to start a broader discussion about the viability or necessity of software testing, as explained by P2: ``\textit{once you have set up the bots, and it is automated, people are less likely to argue about it, which is just a nice effect of bots. Especially for bots that kind of point out failures. I think it's good to have that from bots, and not from people.}'' There are some exceptions, however, when contributors experienced an increase in communication incurred by the bot comments, especially when they do not understand how they might increase the coverage rate. As posed by P9: \textit{In my experience, it causes a longer discussion, because then I have to talk to the engineers like `hey, what kind of a test should I add such as coveralls passes?'''}

\MyPara{Bot feedback pushes contributors to take an action.} Also related to the transparency introduced by the bot comments, and in line with the idea of code review bots as quality gatekeepers, these bots lead developers to take an action: ``\textit{It gives me clear instructions on what I have to do to resolve it. So, I'm very likely to act on it}'' (P2). These bots protect developers from reducing the code's coverage. Therefore, developers would consider either closing the pull request, if it is not worth their time, or following up with the necessary changes: \textit{you have this systematic check that says `okay, that's not good.' And then the developer is saying, `okay, it won't be accepted if I don't provide the test' ''}(P3).

\MyPara{Bot feedback perceived as noise.} Although less recurrent, participants mentioned that in some cases bot comments might be perceived as noise by developers, which disrupts the conversation in the pull request. On the one hand, ``\textit{comments from code coverage bots tend to give you more visibility and provide more context[]}'' (P6). On the other hand, developers complain about the noise these comments introduce to the communication channel. According to P7, the repetitive comments of code coverage bots are ``\textit{disrupting the conversation}'', since  ``\textit{if you have to develop a certain conversation and you have a bot message, this could have a negative impact on the conversation.}'' One of the consequences of this noise incurred by the repetitive bot comments is that ``\textit{[developers] pay less attention to it}'' (P7), impacting the developers communication.

We also asked developers whether they have seen the observed effects on their own projects, and what are the other effects they attribute to the code review bot adoption. The most recurrent (8) observed effect was less communication. As stated by P10: ``\textit{I remember one of the maintainers saying `the tests are missing here.' She always had to post that comment. Then, we adopted the bot to comment on the coverage and had no need for her to comment anymore}.'' Also, 6 participants observed fewer pull requests rejections and faster rejections, and 5 participants have observed more merged pull requests. Finally, developers did not attribute any other effect to the bot introduction.

\MyBox{\textbf{Summary of reasons.} Project maintainers and contributors reported several reasons for more merged pull requests, fewer comments, and fewer and faster rejections. According to them, bot comments help them to understand the state and quality of the contribution, making maintainers more confident to merge pull requests, which also changes the focus of developer discussions.}

\section{Discussion}

Adding a code review bot to a project often represents the desire to enhance feedback about the contributions, helping contributors and maintainers, and achieving improved interpersonal communication, as already discussed by \citet{Storey2016}. Additionally, code review bots can guide contributors toward detecting change effects before maintainers triage the pull requests~\cite{Wessel2018}, ensuring high-quality standards. In this paper, following the study of \citet{Wessel2018}, we focused on monthly activity indicators that are not primarily related to bot adoption, but might be impacted by it. We found that the bot adoption has a statistically significant effect on a variety of activity indicators.

According to the regression results, the monthly number of merged pull requests increased, even faster, after the code review bot adoption. In addition, the number of non-merged pull requests continued to decrease, even faster, after bot adoption. These models showed that after adopting the bot, maintainers started to deal with an increasing influx of contributions ready to be further reviewed and integrated into the codebase. Also, these findings confirm the hypothesis we formulated based on the exploratory case study. According to our participants, the increase in the monthly number of merged pull requests, as well as the decrease in the monthly number of non-merged one, are explained by the transparency introduced by the bot feedback. Contributors started to have faster and clearer feedback on what they needed to do to have their contribution accepted. Further, participants also mentioned that contributors have been pushed to enhance their pull requests based on bot feedback.

In addition, we noticed that just after the adoption of the code review bot the median number of comments slightly increased for merged pull requests. The number of comments on these pull requests could increase due to contributions that drastically reduced the coverage, stimulating discussions between maintainers and contributors. This can happen especially at the beginning of bot adoption, since contributors might be unfamiliar with bot feedback. After that initial period, we found that the median number of comments on merged pull requests decreased each month. According to our participants, less communication could be explained by the transparency and confidence developers gain from bot feedback. Also, developers mentioned that after bot adoption the focus of the developers discussion changed, since there is no need for certain discussions related to coverage. Considering non-merged pull requests, there is no significant change in the number of comments as time passes. These results differ from the case study results, indicating that individual projects reveal different results, which are likely caused by other project-specific characteristics. 

From the regression results, we also noticed an increase in the time spent to merge pull requests just after bot adoption. It makes sense from the contributors' side, since the bot introduces a secondary evaluation step. Especially at the beginning of the adoption, the code review bot might increase the time to merge pull requests due to the need to learn how to meet all bot requirements and obtain a stable code. Maintainers might also deal with an increase in the volume of contributions ready to review and merge, impacting the time spent to review all of them. Further, the regression model shows a decrease in the time spent to review and reject pull requests. Overall, according with our participants it indicates that after the bot adoption maintainers stopped expending effort on pull requests that were not likely to be integrated into the codebase.

As we found in the model of commits on merged pull requests, just after the adoption of the bot the median number of pull request commits increased. The bot provides immediate feedback in terms of proof of failure, which can lead contributors to submit code modifications to change the bot feedback and have their contribution accepted. Overall, the regression models reveal that the monthly number of commits did not change for both merged and non-merged pull requests as time passed. These results differ from the case study results. Nevertheless, even if there is an increase in the number of commits reported in the case study, overall the monthly number of commits are quite stable. For example, for CakePHP it varies from $1$ to $2$ for merged pull requests, and $1$ to $4$ for non-merged pull requests. Additionally, in the main study, we account for control variables, rather than analyzing the monthly number of commits interdependently. As presented in Section \ref{sec:effectsincommits}, for example, the number of comments on pull requests explains the largest amount of variability in these models, indicating that the number of commits is strongly associated with the communication during the pull request review.

\section{Implications and Future Work}

In the following, we discuss implications and future work for researchers and practitioners in light of our results and related literature.

\subsection{Implications for Project Members}

Projects need to make informed decisions on whether to adopt code review bots (or software bots in general) and how to use them effectively. We found that the dynamics of pull requests changed following the adoption of code review bots. Hence, besides understanding the effects on code quality, practitioners and open-source developers should become aware of other consequences of bot adoption and take countermeasures to avoid the undesired ones. For example, our statistical findings show a decrease in the amount of discussion between humans after the bot's introduction. According to developers, this effect is likely to be explained by more visibility and transparency, or the changes in the focus of the discussions. However, developers might also perceive bot comments as noise, which disrupts the conversation in the pull request. Thus, project members should be aware of these possible side effects since noise is a recurrent problem when adopting bots on pull requests~\cite{Wessel2021CSCW}. For instance, they might consider re-configuring the bot to avoid some behaviors, such as high frequency of actions---bots performing repetitive actions, such as creating numerous pull requests and leaving dozen of comments in a row---and comments verbosity---bots providing comments with dense information.

\subsection{Implications for Researchers}

For researchers interested in software bots, it is important to understand the role of code review bots in the bot landscape. It is important to understand how such bots affect the interplay of developers in their effort to develop software, and our study provides the first step in this direction. Considering that bot output is mostly text-based, how bots present content can highly impact developers' perceptions~\cite{botmit,chaves2020should}. Additional effort is necessary to investigate how the developers' cognitive styles~\cite{gendermag,gendermag2} might influence the way developers interpret the bot comments' content. In this way, future research can investigate how people with different cognitive styles handle bot messages and learn from them. Other social characteristics of the bots can also be investigated in this context~\cite{chaves2020should}. Future research can lead to a set of guidelines on how to design effective messages for different cognitive styles and developer profiles. Further, developers complain about the information overload caused by repetitive bot behavior on pull requests, which has received some attention from the research community~\cite{Wessel2018,Wessel2020,erlenhov2020empirical}, but remains a challenging problem. In fact, there is room for improvement on human-bot collaboration on social coding platforms. When they are overloaded with information, teams must adapt and change their communication behavior~\cite{ellwart2015managing}. Therefore, there is also an opportunity to investigate changes in developers' behavior imposed by the effects of information overload. Additional research can also investigate how to use code reviews bots to support the training of new software engineers~\cite{pinto2017training}.

Previous work by \citet{Wessel2018} has already mentioned that bot support for newcomers is both challenging and desirable. In a subsequent study, \citet{Wessel2020whatexpect} reported that although bots could make it easier for some newcomers to submit a high-quality pull request, bots can also provide newcomers with information that can lead to rework, discussion, and ultimately dropping out from contributing. It is reasonable to expect that newcomers who receive friendly feedback will have a higher engagement level and thus sustain their participation on the project. Hence, future research can help bot designers by providing guidelines and insights on how to support new contributors. Additional effort is also necessary to investigate the impact of code review bots' feedback for newcomers, who already face a variety of barriers~\cite{balali2018newcomers,steinmacher2015social}.

\subsection{Implications for Code Review Bots}

To avoid side effects of using code review bots, such as noise, bots should provide mechanisms to enable better configurable control over their actions, rather than just turn off bot comments. It is important to have easy mechanisms so project maintainers can turn off or pause a bot at any time. Further, these mechanisms need to be explicitly announced during bot adoption (e.g., noiseless configuration, preset levels of information). It is essential to provide a more flexible way for bots to interact, incorporating rich user interface elements to better engage users.

\section{Related work}

In this section, we describe the studies related to the usage and impact of software bots. Further, we summarize works that employed regression discontinuity design (RDD) to account for the intervention effects on software development activities on GitHub.

\subsection{Software Bots on Social Coding Platforms}

Software bots are software applications that integrate their work with human tasks, serving as interfaces between users and other tools~\cite{Storey2017,DBLP:journals/corr/LebeufSZ17}, and providing additional value to human users~\cite{Lebeuf2019}. Software bots frequently reside on platforms where users work and interact with other users~\cite{lebeuf2018software}. On the GitHub platform, bots have user profiles to interact with the developers, executing well-defined tasks~\cite{Wessel2018}. 

Bots support social and technical activities in software engineering, including communication and decision-making~\cite{Storey2016}. 
Bots are particularly relevant in social-coding platforms~\cite{Dabbish2012}, such as GitHub, where the pull-based model~\cite{gousios2014exploratory} offers several opportunities for community engagement, but at the same time increases the workload for maintainers~\cite{Gousios2016, pinto2016more}. Open-source communities have been adopting bots to reduce the workload with a variety of automated repetitive tasks on GitHub pull requests~\cite{Wessel2018}, including repairing bugs~\cite{urli2018design,Monperrus2019}, refactoring source code~\cite{Wyrich2019}, recommending tools~\cite{Brown2019}, updating dependencies~\cite{mirhosseini2017can}, fixing static analysis violations~\cite{botc3pr,Serban2021}, suggesting code improvements~\cite{phan2020teddy}, and predicting defects~\cite{Khanan2020}. 

\citet{Storey2016} and Paikari and van der Hoek~\cite{Paikari.vanDerHoek:2018} highlight that the potentially negative impact of task automation through bots is being overlooked. \citet{Storey2016} claim that bots are often used to avoid interruptions to developers' work, but may lead to other, less obvious distractions. While previous studies provide recommendations on how to develop bots and evaluate bots' capabilities and performance, they do not draw attention to the impact of bot adoption on software development or how software engineers perceive the bots' impact. Since bots are seen as new team members~\cite{Monperrus2019}, we expected that bots would impact group dynamics in a way that differs from non-bot forms of automation.

\citet{Wessel2018} investigated the usage and impact of software bots to support contributors and maintainers with pull requests. After identifying bots on popular GitHub repositories, the authors classified them into 13 categories according to their tasks. Unlike \citet{Wessel2018}, we focused on understanding the effects of a specific bot type, which is the most frequently used category of bots. 
In a preliminary study, \citet{Wessel2020whatexpect} surveyed 127 open source maintainers experienced in using code review bots. While maintainers report that bots satisfy their expectations regarding enhancing developers' feedback, reducing maintenance burden, and enforcing code coverage, they also perceived unexpected effects of having a bot, including communication noise, more time spent with tests, and  newcomers' dropout. Our work extends this preliminary investigation by combining analysis of GitHub data with semi-structured interviews conducted with open-source developers. This study looks at how bots change the pull request dynamics and its reasons from practitioners' perspectives.

\subsection{Using RDD to Access the Effects of Interventions on Software Development}

\begin{landscape}
\centering
\begin{table}[]
\scriptsize
\scalefont{0.7}
\caption{Literature review related to Software development and RDD on GitHub}
\label{tab:comparing-works}
\begin{threeparttable}
\begin{tabular}{lccrllrcccc}
\toprule
\multicolumn{1}{c}{}  &     & \multicolumn{9}{c}{\textbf{Common variables}}               \\ \cmidrule{3-11} 
\multicolumn{1}{l}{\multirow{-2}{*}{\textbf{Study}}} & \multirow{-2}{*}{\textbf{Intervention}} & \multicolumn{2}{c}{\textbf{Comments}} & \multicolumn{2}{c}{\textbf{Commits}}   & \multicolumn{1}{c}{\textbf{Issues}}    & \multicolumn{2}{c}{\textbf{PR Latency}}  & \multicolumn{2}{c}{\textbf{Pull Requests}} \\ \midrule
\citet{zhao2017impact}   & \cellcolor[HTML]{ECF4FF}\begin{tabular}[c]{@{}c@{}}\textbf{Travis CI}\\ (adoption)\end{tabular} & \cellcolor[HTML]{EFEFEF}\textbf{}   & \multicolumn{1}{c}{\cellcolor[HTML]{EFEFEF}\textbf{}} &  \multicolumn{2}{c}{\cellcolor[HTML]{CBFFCA}\textbf{Merge commits}}  & \cellcolor[HTML]{EFEFEF}                                           & \multicolumn{2}{c}{\cellcolor[HTML]{CBFFCA}\textbf{\begin{tabular}[c]{@{}c@{}}Time to \\ close PRs\end{tabular}}} & \multicolumn{2}{c}{\cellcolor[HTML]{CBFFCA}\textbf{Closed PRs}} \\\hline
\citet{cassee2020silent} & \cellcolor[HTML]{ECF4FF}\begin{tabular}[c]{@{}c@{}}\textbf{Travis CI}\\ (adoption)\end{tabular}          & \cellcolor[HTML]{FFCCC9}\textbf{\begin{tabular}[c]{@{}c@{}}PR \\ comments\end{tabular}}           & \multicolumn{1}{c}{\cellcolor[HTML]{FFCCC9}\textbf{\begin{tabular}[c]{@{}c@{}}PR review \\ comments\end{tabular}}}         & \multicolumn{2}{c}{\cellcolor[HTML]{FFFFC7}\textbf{\begin{tabular}[c]{@{}c@{}}Commits after \\ create the PR\end{tabular}}}         & \cellcolor[HTML]{EFEFEF}                                           & \cellcolor[HTML]{EFEFEF}                                                                                           & \cellcolor[HTML]{EFEFEF}  & \cellcolor[HTML]{EFEFEF}\textbf{} & \cellcolor[HTML]{EFEFEF}\textbf{}\\\hline
\citet{guo2019studying} & \cellcolor[HTML]{ECF4FF}\begin{tabular}[c]{@{}c@{}}\textbf{Travis CI}\\ (adoption)\end{tabular}          & \multicolumn{1}{l}{\cellcolor[HTML]{EFEFEF}}                                                      & \cellcolor[HTML]{EFEFEF}                                                                                                   & \multicolumn{2}{l}{\cellcolor[HTML]{EFEFEF}}                                                                                                            & \cellcolor[HTML]{EFEFEF}                                           & \multicolumn{2}{c}{\cellcolor[HTML]{FFFFC7}\textbf{\begin{tabular}[c]{@{}c@{}}Time to \\ deliver PRs\end{tabular}}}  & \cellcolor[HTML]{EFEFEF}\textbf{} & \cellcolor[HTML]{EFEFEF}\textbf{} \\\hline
\citet{kavaler2019tool} & \begin{tabular}[c]{@{}c@{}}\textbf{Quality} \\ \textbf{assurance} \\ \textbf{tools} \\ (adoption)\end{tabular}             & \multicolumn{1}{l}{\cellcolor[HTML]{EFEFEF}}                                                      & \cellcolor[HTML]{EFEFEF}  & \multicolumn{2}{l}{\cellcolor[HTML]{EFEFEF}} & \multicolumn{1}{c}{\cellcolor[HTML]{FFCCC9}\textbf{Opened}} & \cellcolor[HTML]{EFEFEF}                                                                                           & \cellcolor[HTML]{EFEFEF}  & \cellcolor[HTML]{EFEFEF}\textbf{} & \cellcolor[HTML]{EFEFEF}\textbf{}   \\\hline
\citet{wessel2020effects} & \begin{tabular}[c]{@{}c@{}}\textbf{Code review} \\ \textbf{bots}\\ (adoption)\end{tabular}                        & \cellcolor[HTML]{FFCCC9}\textbf{\begin{tabular}[c]{@{}c@{}}Merged \\PRs\end{tabular}} & \multicolumn{1}{c}{\cellcolor[HTML]{FFFFC7}\textbf{\begin{tabular}[c]{@{}c@{}}Non-merged \\PRs\end{tabular}}} & \multicolumn{1}{c}{\cellcolor[HTML]{FFFFC7}\textbf{\begin{tabular}[c]{@{}c@{}}Merged\\PRs\end{tabular}}} & \multicolumn{1}{c}{\cellcolor[HTML]{FFFFC7}\textbf{\begin{tabular}[c]{@{}c@{}}Non-merged\\PRs\end{tabular}}} & \cellcolor[HTML]{EFEFEF}                                           & \multicolumn{1}{c}{\cellcolor[HTML]{FFCCC9}\textbf{\begin{tabular}[c]{@{}c@{}}Time to \\ reject PRs\end{tabular}}} & \cellcolor[HTML]{FFFFC7}\textbf{\begin{tabular}[c]{@{}l@{}}Time to \\ merge PRs\end{tabular}} & \cellcolor[HTML]{CBFFCA}\textbf{\begin{tabular}[c]{@{}l@{}}Merged\\ PRs\end{tabular}} & \cellcolor[HTML]{FFCCC9}\textbf{\begin{tabular}[c]{@{}l@{}}Non-merged\\PRs\end{tabular}}\\\hline
\citet{kinsman2021} & \begin{tabular}[c]{@{}c@{}}\textbf{Code review} \\ \textbf{bots}\\ (adoption)\end{tabular}                        & \cellcolor[HTML]{FFFFC7}\textbf{\begin{tabular}[c]{@{}c@{}}Merged \\ PRs\end{tabular}} & \multicolumn{1}{c}{\cellcolor[HTML]{FFFFC7}\textbf{\begin{tabular}[c]{@{}c@{}}Non-merged \\PRs\end{tabular}}} & \multicolumn{1}{c}{\cellcolor[HTML]{FFCCC9}\textbf{\begin{tabular}[c]{@{}c@{}}Merged\\PRs\end{tabular}}} & \multicolumn{1}{c}{\cellcolor[HTML]{FFFFC7}\textbf{\begin{tabular}[c]{@{}c@{}}Non-merged\\PRs\end{tabular}}} & \cellcolor[HTML]{EFEFEF}                                           & \multicolumn{1}{c}{\cellcolor[HTML]{FFFFC7}\textbf{\begin{tabular}[c]{@{}c@{}}Time to \\ reject PRs\end{tabular}}} & \cellcolor[HTML]{FFFFC7}\textbf{\begin{tabular}[c]{@{}l@{}}Time to \\ merge PRs\end{tabular}} & \multicolumn{1}{c}{\cellcolor[HTML]{FFFFC7}\textbf{\begin{tabular}[c]{@{}l@{}}Merged\\ PRs\end{tabular}}} & \multicolumn{1}{c}{\cellcolor[HTML]{CBFFCA}\textbf{\begin{tabular}[c]{@{}l@{}}Non-merged\\PRs\end{tabular}}}\\ \hline
\citet{trockman2018adding} & \begin{tabular}[c]{@{}c@{}}\textbf{Repository} \\ \textbf{badges}\\ (adoption)\end{tabular}                       & \multicolumn{1}{l}{\cellcolor[HTML]{EFEFEF}}                                                      & \cellcolor[HTML]{EFEFEF}                                                                                                   & \multicolumn{2}{l}{\cellcolor[HTML]{EFEFEF}}                                                                                                          & \cellcolor[HTML]{EFEFEF}                                           & \cellcolor[HTML]{EFEFEF}                                                                                           & \cellcolor[HTML]{EFEFEF}                                                                      & \cellcolor[HTML]{EFEFEF}\textbf{} & \cellcolor[HTML]{EFEFEF}\textbf{} \\ \hline
\citet{zimmermann2019impact} & \begin{tabular}[c]{@{}c@{}}\textbf{Bug tracker} \\ (move from \\ Bugzilla to \\ GitHub)\end{tabular}     & \multicolumn{2}{c}{\cellcolor[HTML]{CBFFCA}\textbf{\begin{tabular}[c]{@{}c@{}}Bug tracker/\\ Issue comments\end{tabular}}} & \multicolumn{2}{l}{\cellcolor[HTML]{EFEFEF}}  & \multicolumn{1}{c}{\cellcolor[HTML]{CBFFCA}\textbf{Opened}} & \cellcolor[HTML]{EFEFEF}                                                                                           & \cellcolor[HTML]{EFEFEF}  & \cellcolor[HTML]{EFEFEF}\textbf{} & \cellcolor[HTML]{EFEFEF}\textbf{} \\\hline
\citet{moldon2020gamification}  & \begin{tabular}[c]{@{}c@{}}\textbf{Gamification} \\ \textbf{mechanisms} \\ (removing \\ from GitHub)\end{tabular} & \multicolumn{1}{l}{\cellcolor[HTML]{EFEFEF}}                                                      & \cellcolor[HTML]{EFEFEF}                                                                                                   & \multicolumn{2}{l}{\cellcolor[HTML]{EFEFEF}}                                                                                                             & \cellcolor[HTML]{EFEFEF}                                           & \cellcolor[HTML]{EFEFEF}                                                                                           & \cellcolor[HTML]{EFEFEF} & \cellcolor[HTML]{EFEFEF}\textbf{} & \cellcolor[HTML]{EFEFEF}\textbf{} \\ \hline \citet{walden2020impact} & \begin{tabular}[c]{@{}c@{}}\textbf{Major Security}\\ \textbf{Event}\\ (Heartbleed)\end{tabular}                   & \multicolumn{1}{l}{\cellcolor[HTML]{EFEFEF}}                                                      & \cellcolor[HTML]{EFEFEF}                                                                                                   &   \multicolumn{2}{c}{\cellcolor[HTML]{CBFFCA}\textbf{Merge commits}}        & \cellcolor[HTML]{EFEFEF}                                    & \cellcolor[HTML]{EFEFEF}                                                                                           & \cellcolor[HTML]{EFEFEF} & \cellcolor[HTML]{EFEFEF}\textbf{} & \cellcolor[HTML]{EFEFEF}\textbf{} \\ \bottomrule                                                                    
\end{tabular}
\begin{tablenotes}
 \item \textbf{Legend:} \begin{tabular}[c]{@{}l@{}}\cellcolor[HTML]{CBFFCA} \textbf{Increase}\end{tabular} \begin{tabular}[c]{@{}l@{}}\cellcolor[HTML]{FFCCC9} \textbf{Decrease}\end{tabular} \begin{tabular}[c]{@{}l@{}}\cellcolor[HTML]{FFFFC7} \textbf{Does not change}\end{tabular}
\end{tablenotes}
\end{threeparttable}
\end{table}
\end{landscape}

In the software engineering domain, several researchers have been applying \textit{Regression Discontinuity Design} (RDD) to model the effects of a variety of interventions on development activities over time. To understand the similarities between those studies, we conducted an extensive search for empirical works that employed RDD to investigate interventions in software development on GitHub in general. In Table~\ref{tab:comparing-works} we summarize these studies, presenting an overview of what interventions have been used (e.g., bots, CI), what dependent variables have been studied, and what results have been obtained.

\citet{zhao2017impact} introduced the RDD usage to study software development activities. \citet{zhao2017impact} focused on the impact of the Continuous Integration  (CI) tool's introduction on development practices. Conducting the statistical analysis on GitHub repositories, they found that adopting Travis CI leads to an increase in the number of merge commits, number of closed pull requests, and in pull request latency. With these results they also confirm earlier results about the benefits of CI, such as a better adherence to best practices.
Meanwhile, \citet{cassee2020silent} studied the effects of Travis CI on conserving developers' efforts during code review. Analyzing the pull requests' general comments and the review comments, which are associated with specific lines of code on the pull request, they found that the communication decreased after the CI adoption. At the same time, the trends in the commits after the creation of the pull requests remained unaffected.
Also regarding CI, \citet{guo2019studying} investigated the impact of its adoption on the delivery time of pull requests. They find no evidence of CI affecting the pull request delivery time in the studied projects.

In addition to the studies of CI, prior work has also investigated the impact of other automation tools designed to support developers during code review or while performing other repetitive tasks on pull requests. \citet{kavaler2019tool}, for example, investigated the impact of linters, dependency managers, and coverage reporter tools on GitHub projects across time. The results of applying RDD showed that tools are associated with a decrease in the monthly number of opened issues.
\citet{trockman2018adding} explored the impacts of the usage badges on GitHub repositories. They found that badges displaying the build status, test coverage, and up-to-dateness of dependencies are associated with more tests, more quality pull requests, and fresher dependencies. 
\citet{kinsman2021} studied the effect of GitHub Action adoption by GitHub projects. The results revealed that introducing a GitHub Action leads to an increase in the number of rejected pull requests and a decrease in the commits in the merged pull requests. This differs from our results, which might be explained by the variety of tasks performed by the GitHub Actions in the study, and consequently their impacts on pull request activities. 

Other studies have been investigating interventions that are not related to a tool adoption. For example, \citet{zimmermann2019impact} investigated the impact of switching from one bug tracker to another. They found that the switch induces an increase in issue reporting, particularly by the project core developers. Moreover, when moving from Bugzilla to GitHub, the communication between maintainers and contributors in the issues also increased.
\citet{moldon2020gamification} focused on how developers' behavior was impacted by the removal of the daily activity streak counters from the user profile. The results show that the developer activity decreased on weekends compared to weekdays. According to the authors, the activity counters were influencing developers to contribute on days they would have otherwise rested.
\citet{walden2020impact} employed RDD to assess the impact of a major security event on the evolution of a specific project called OpenSSL. As a result of the intervention, the number of monthly commits increased and the code complexity decreased.

In short, we showed an overview of how RDD have been used in empirical software engineering studies. As described in the Table~\ref{tab:comparing-works}, previous works investigated distinct variables. Even selecting related variables as ``comments'', each study focused on different types of comments (e.g. general pull requests comments, review comments, issue comments), or comments applied to different scenarios (e.g. comments on merged and non-merged pull requests).
Our work extends this literature by providing a more in-depth investigation of the effects of a specific type of automation, namely code review bot adoption.

\section{Limitations and Threats to Validity}

In this section, we discuss the limitations and potential threats to validity of our study, their potential impact on the results, and how we have mitigated them~\cite{wohlin2012experimentation}.

\MyPara{External Validity:} While our results only apply to OSS projects hosted on GitHub, many relevant projects are currently hosted on this platform~\cite{Dias2016}. Our selection of projects also limits our results. Therefore, even though we considered a large number of projects and our results indicate general trends, we recommend running segmented analyses when applying our results to a given project. For replication purposes, we made our data and source code publicly available.\footnote{https://doi.org/10.5281/zenodo.4618498}

\MyPara{Construct Validity:} One of the constructs in our study is the ``first bot comment on a pull request'' as a proxy to the ``time of bot adoption'' on a project. A more precise definition of this adoption time would have involved the integration date, which is not provided by the GitHub API. 
Moreover, recent studies have observed `mixed’ GitHub accounts, i.e., accounts shared by a human and a bot~\cite{GolzadehDCM21,CasseeKCS21}, \emph{e.g.}, exhibiting user name and avatar and posting both human-written and bot-generated comments. A more precise definition of bot adoption should consider activity of the `mixed' accounts as well.
Hence, the validity of the ``time of bot adoption'' construct might have been threatened by the definition. We reduce this threat by excluding the period of $15$ days immediately before and after adoption from all analyses.
Moreover, \citet{Kalliamvakou2014} stated that many merged pull requests appear non-merged, which could also affect the construct validity of our study, since we consider the number of merged pull requests. 
To increase construct validity and improve the reliability of our qualitative findings, we employed a constant comparison method~\cite{glaser2017discovery}. In this method, each interpretation is constantly compared with existing findings as it emerges from the qualitative analysis.

\MyPara{Internal Validity:} To reduce internal threats, we applied multiple data filtering steps to the statistical models. To confirm the robustness of our models, we varied the data filtering criteria, for example, by filtering projects that did not receive pull requests in all months, instead of at least $6$ months, and observed similar phenomena. Projects that disabled the bot during the period we considered might be a threat. However, detecting whether a project disabled the bot or not is challenging. The GitHub API does not provide this information. We reduce this threat by removing from our dataset projects without bot comments during the last four months of analysis. Additionally, we added several controls that might influence the independent variables to reduce confounding factors. However, in addition to the already identified dependent variables, there might be other factors that influence the activities related to pull requests. These factors could include the adoption of other code review bots, or even other types of bots and non-bot automation. To treat this, we removed projects that adopted more than one bot, based on the list of bots provided by \citet{Wessel2018}.
To ensure information saturation, we continued recruiting participants and conducting interviews until we came to an agreement that no new significant information was found. As posed by Strauss and Corbin~\cite{strauss1997grounded}, sampling should be discontinued once the collected data is considered sufficiently dense and data collection no longer generates new information.

\section{Conclusion}

In this work, we conducted an exploratory empirical investigation of the effects of adopting bots to support the code review process on pull requests. While several code review bots have been proposed and adopted by the OSS community, relatively little has been done to evaluate the state of practice. To understand the impact on practice, we statistically analyzed data from $1,194$ open source projects hosted on GitHub. Further, we had a deep investigation into the reasons of the identified impacts. We interviewed 12 project maintainers and contributors experienced with code review bots.

By modeling the data around the introduction of a code review bot, we notice different results from merged pull requests and non-merged ones. We see that the monthly number of merged pull requests of a project increases after the adoption of a code review bot, requiring less communication between maintainers and contributors. At the same time, code review bots can lead projects to reject fewer pull requests. Afterwards, when interviewing developers we found a comprehensive set of reasons for these effects. First of all, bot comments help contributors and maintainers to be aware the state and quality of the contribution, making maintainers more confident to merge pull requests, which also changes the focus of developers' discussions.

Practitioners and open-source maintainers may use our results to understand how group dynamics can be affected by the introduction of a code review bot, and to design counter-measurements to avoid undesired effects. Future work includes a qualitative investigation of the effects of adopting a bot and the expansion of our analysis for other types of bots, activity indicators, social coding platforms, and statistical approaches, such as counterfactual time series~\cite{MurphyHill2019}.

\begin{acknowledgements}
This work was partially supported by the Coordenação de Aperfeiçoamento de Pessoal de Nível Superior – Brasil (CAPES) – Finance Code 001, CNPq (grant 141222/2018-2), and National Science Foundation (grants 1815503 and 1900903).
We also thank the Open Source developers who spent their time to participate in our research.
\end{acknowledgements}


%
%

\bibliographystyle{spbasic}      
\bibliography{main}   

\end{document}